\documentclass[10pt,journal,epsfig,twoside]{IEEEtran}
\usepackage{amsmath}
\usepackage{graphicx} 
\usepackage{epstopdf}
\usepackage{amssymb}
\usepackage{amsfonts}
\usepackage{amsthm}
\usepackage{cite}
\usepackage{bm,comment}
\usepackage{algorithm}
\usepackage{algpseudocode}

\usepackage{subeqnarray}
\usepackage{subfigure}
\usepackage{multicol}
\usepackage{multirow}
\usepackage{diagbox}
\usepackage{slashbox}
\usepackage{stfloats}
\usepackage{float}
\usepackage{color} 
\usepackage{cases}

\newtheorem{proposition}{Proposition}

\newcommand{\tabincell}[2]{\begin{tabular}{@{}#1@{}}#2\end{tabular}} 
\renewcommand{\baselinestretch}{0.97}

\begin{document}
\title{A Unified 3D Beam Training and Tracking Procedure for Terahertz Communication}
\author{Boyu Ning, Zhi Chen, \IEEEmembership{Senior Member, IEEE}, Zhongbao Tian, \\ Chong Han, \IEEEmembership{Member, IEEE}, and Shaoqian Li, \IEEEmembership{Fellow, IEEE}
\vspace{-6pt}
\thanks{ 
This work was presented in part at IEEE International Conference on Communications (ICC), 2021\cite{icc}, https://doi.org/10.1109/ICCWorkshops50388.2021.9473577. This work was supported in part by the National Key R$\&$D Program of China under Grant 2018YFB1801500.
This work was also supported in part by National Natural Science Foundation of China (NSFC) under Grant No. 62171280.}
\thanks{B. Ning, Z. Chen, Z. Tian, and S. Li are with the National Key Laboratory of Science and Technology on Communications, University of Electronic Science and Technology of China, Chengdu 611731, China (e-mails: boydning@outlook.com; chenzhi@uestc.edu.cn; vincent11231@outlook.com; lsq@uestc.edu.cn).}
\thanks{C. Han is with the Terahertz Wireless Communications (TWC) Laboratory, Shanghai Jiao Tong University, Shanghai 200240, China (e-mail: chong.han@sjtu.edu.cn).}}
\maketitle

\begin{abstract}
Terahertz (THz) communication is considered as an attractive way to overcome the bandwidth bottleneck and satisfy the ever-increasing capacity demand in the future. Due to the high directivity and propagation loss of THz waves, a massive MIMO system using beamforming is envisioned as a promising technology in THz communication to realize high-gain and directional transmission. However, pilots, which are the fundamentals for many beamforming schemes, are challenging to be accurately detected in the THz band owing to the severe propagation loss. In this paper, a unified 3D beam training and tracking procedure is proposed to effectively realize the beamforming in THz communications, by considering the line-of-sight (LoS) propagation. In particular, a novel quadruple-uniform planar array (QUPA) architecture is analyzed to enlarge the signal coverage, increase the beam gain, and reduce the beam squint loss. Then, a new 3D grid-based (GB) beam training is developed with low complexity, including the design of the 3D codebook and training protocol. Finally, a simple yet effective grid-based hybrid (GBH) beam tracking is investigated to support THz beamforming in an efficient manner. The communication framework based on this procedure can dynamically trigger beam training/tracking depending on the real-time quality of service. Numerical results are presented to demonstrate the superiority of our proposed beam training and tracking over the benchmark methods.
\end{abstract}
\IEEEpeerreviewmaketitle

\section{Introduction}
Terahertz (THz) communication is considered as a key wireless technology to alleviate the spectrum bottleneck and support high data rates in the future\cite{tt1}. The THz band, ranging from $0.1$ to $10$ THz, supports huge transmission bandwidth, and owns multiple appealing transmission windows separated by the attenuation absorption peaks\cite{tt2}. Despite huge bandwidth on unlicensed spectrum, the challenge of using THz band spectrum comes from the severe propagation loss due to both spreading path loss and molecular absorption\cite{tt3}.

To compensate for the propagation loss, various technologies, e.g., massive multiple-input multiple-output (MIMO) \cite{m1}, coordinated multi-point transmission \cite{cm}, and intelligent reflecting surface\cite{b1}, can be integrated in THz communications to provide effective spatial diversity gains. In the THz massive MIMO systems as we concern, the transmitter and receiver equipped with large-scale antenna arrays can realize directional communication with sufficient beam gains by dynamically controlling the amplitude and phase shifts on each antenna element\cite{mm}. Nevertheless, the conventional beamforming technologies usually require accurate channel state information (CSI) between the transmitter and receiver for optimizing data transmission, which is challenging in THz systems since the pilot signals, generally being transmitted without adequate beam gains, may not be effectively detected by the receiver owing to the severe propagation loss\cite{b2}. 

This issue has already been encountered in millimeter-wave (mmWave) systems. In this context, a new approach, called \emph{beam training}, has been proposed for effective directional communication by testing beam pairs without requiring any CSI\cite{t1,t2,t3,ywang,Qsu,ones,parallel,wzhong,thzh}. A feasible beam training scheme should contain the designs of codebook and training protocol\cite{wide1,wide2,wide3,wide4,wide5}, in which the former determines the radiation pattern of the beams (i.e., codewords), while the latter focuses on how to use these predefined beams to realize beam alignment at the transmitter and receiver. After a successful beam alignment, beam tracking technologies can be applied for mobile transceivers, which assist to reduce the training overhead\cite{tr1,tr2,tr3,mk1,mk2,mk3,tr4,tr5}. It is worth noting that the existing beam tracking techniques are developed independently of the beam training techniques, which may rely on a certain antenna geometry, transceiver architecture, as well as form of channel information. To facilitate a generic system design, a unified beam tracking and training design is thus stringently needed. Besides, most existing schemes are tailored for the uniform linear array, i.e., 2D beamforming. However, due to the high directivity of THz wave, 3D beamforming by uniform planar array (UPA) has practical potential for emerging THz applications, e.g., efficient integrated networks of terrestrial links, unmanned aerial vehicles (UAV), and satellite communication systems\cite{wm,sk}.

To this end, we propose a unified 3D beam training and tracking procedure for THz communications in this paper. As a holistic design, this procedure contains many novel aspects, in terms of architecture, framework, 3D codebook, and training/tracking protocol. In particular, this procedure only needs to search the codewords according to our proposed protocol, instead of calculating the real-time beamforming according to the CSI, thus facilitating the low-complexity implementation of beam training and tracking in THz communications in practice. It is worth mentioning that our proposed scheme is catering to the line-of-sight (LoS) propagation path, which is the dominant component of the THz channel. When it is applied to the multi-path lower-frequency channel, e.g., in mmWave indoor scenario, the efficiency may be compromised since the received signals would be interfered by the non-negligible non-line-of-sight (NLoS) components. The contributions of this paper are summarized as follows.
\begin{itemize}
\item We consider a novel quadruple-uniform planar array (QUPA) architecture that covers omni-direction in the azimuth and $\frac{\pi }{2}$ range in the elevation domains, in which each UPA only supports $\pm\frac{\pi }{4}$ range three-dimensionally. Compared to the conventional single UPA architecture that covers $\pm\frac{\pi }{2}$ range in both azimuth and elevation, each UPA in the QUPA has substantially less angular deflection and the beam squint loss can be reduced. Besides, since each UPA only serves a confined range, higher array gains can be achieved by using the directional antenna element tailored for the certain coverage.

\item  We propose a holistic communication framework to build a unified 3D beam training and tracking procedure. Instead of performing beam training or tracking over a fixed frequency, our proposed communication framework adopts dynamic on-demand beam training/tracking depending on the real-time quality of service. This can effectively reduce potential outages that may occur in the fixed-frequency-based conventional schemes, owing to the narrow-beam transmission and fast movement of transceivers in the THz communications.


\item  For realizing beam training, we first develop a new 3D hierarchical codebook that pre-defines some codewords for narrow beams and wide beams stage-by-stage.  Although the 3D beams can be simply written as the Kronecker product of the beams the 2D codebook\cite{wide5}, this approach yields an irregular beams' coverage since the beams' azimuth distribution are various at different elevation angles\footnote{For our considered QUPA architecture, if straightforwardly using the Kronecker product of existing 2D codewords, the azimuth coverage expands when beams are above/below 90 degrees of elevation angle, which makes the total coverage of the QUPA cannot constitute an exact sphere.}. By contrast, our proposed approach specifies how to judiciously design the distribution of beams within a given 3D coverage requirement, which guarantees the maximum worst-case training performance. Then, we develop a new 3D training protocol to find the optimal narrow-beam pair based on our proposed codebook, which incurs significantly lower training complexity compared to the existing schemes. The codebook and the protocol are developed based on a 3D grid, and we call this scheme grid-based (GB) training.

\item  For realizing beam tracking, we develop an efficient protocol that searches the codewords in a fast and efficient manner, rather than calculating the channel variations by, e.g., location-based prediction, angular-based prediction, and Kalman filters, in conventional schemes with high complexity. The proposed protocol combines two tracking modes with different search times. The first mode needs to search the beams in the vicinity of the formerly used beam pair on our predefined grid, while the second one directly chooses a new beam pair for connection based on the changing trend of the previously used beam pairs on the grid. As there are two tracking modes jointly realizing the beam alignment, we call this scheme a grid-based hybrid (GBH) tracking.

\item  Numerical results demonstrate the superiority of our proposed beam training and tracking over the benchmark methods. Compared to the existing training codebooks, our proposed wide beams have a smaller dead zone with the lowest misalignment probability during the training, while our proposed narrow beams show no overlap between different UPAs and yield the highest received SNR after the training. Compared to the existing tracking schemes, our proposed beam tracking yields the highest worst-case performance. By combining the first and second tracking modes, no outage occurs via our proposed beam tracking over all the test time. 

\end{itemize}

The rest of this paper is organized as follows. Section II introduces the system and describes the problem. In Section III, we present the framework of our unified beam training and tracking procedure. Section IV develops the beam training and tracking approaches. Section V demonstrates the performance improvement of the proposed scheme through the numerical results. Finally, we conclude the paper in Section VI.

\indent \emph{Notation:} We use small normal face for scalars, small bold face for vectors, and capital bold face for matrices. The superscript ${{\rm{\{ }} \cdot \}^T}$ and ${{\rm{\{ }} \cdot \}^H}$ denote the transpose and Hermitian transpose, respectively. $\mathcal{CN}(\mu,\sigma^2)$ means circularly symmetric complex Gaussian (CSCG) distribution with mean of $\mu$ and variance of $\sigma^2$. $|\cdot|$ represents the modulus operator. $\bmod _N (i)$ returns the remainder after division of $i$ by $N$. ${\rm{ceil}}(\cdot)$ returns the nearest integer greater than or equal to its argument.

\section{System and Problem Descriptions}
In this section, we introduce the considered system model and formulate the problems of beam training and tracking.
\begin{figure} 
    \centering 
  \includegraphics[width=3in]{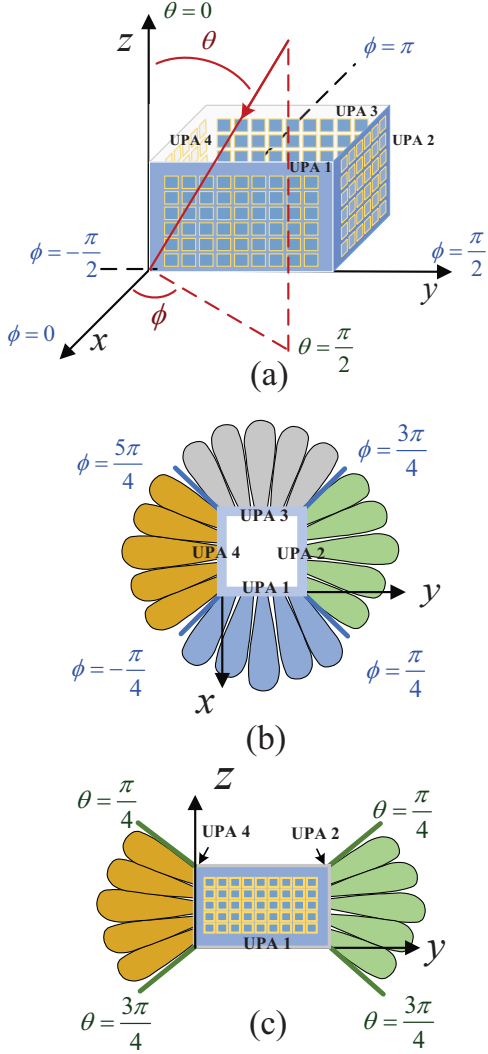}
\caption{(a) QUPA geometry. (b) Transmit and receive ranges of QUPA on the $xy$-plane. (c) Transmit and receive ranges of QUPA on the $yz$-plane.}\label{mo}
\vspace{-12pt}
\end{figure}
\subsection{System Model}
We consider a point-to-point THz massive MIMO system with four half-wave spaced UPAs, i.e., QUPA, equipped at the transmitter and receiver, respectively. Without loss of generality, we assume that both the transmitter and receiver have the same architecture where four identical UPAs with $N_a$ elements are equipped around a cube. As shown in Fig. \ref{mo}(a), we use $x$, $y$, and $z$-axes to refer to the axes of the standard Cartesian coordinate system for the QUPA. In the case of the first UPA, with $N_y$ and $N_z$ elements on the $y$ and $z$-axes respectively ($N_a=N_y N_z$), the array response vector can be expressed in a conventional form\footnote{We assume the signal phase at the center of the UPA is zero.}, i.e.,
\begin{equation}
\begin{split}
&{{\bf{a}}_1}(\phi ,\theta ) \!=\! \frac{1}{{\sqrt {{N_a}} }}[{e^{j\pi [ - \frac{{({N_y} - 1)}}{2}\sin \phi \sin \theta  - \frac{{({N_z} - 1)}}{2}\cos \theta ]}},...,\\
&\;\;\;\;\;\;\;\;\;\;\;\;\;\;\;\;\;\;\qquad{e^{j\pi ({n_y}\sin \phi \sin \theta  + {n_z}\cos \theta )}},...,\\
&\;\;\;\;\;\;\;\;\;\;\;\;\;\;\;\;\;\qquad\qquad{e^{j\pi [\frac{{({N_y} - 1)}}{2}\sin \phi \sin \theta  + \frac{{({N_z} - 1)}}{2}\cos \theta ]}}{]^T},
\end{split}
\end{equation} 
where $\phi$ and $\theta$ are the azimuth angle to $x$-axis and the elevation angle to $z$-axis respectively, ${n_y} =  - \frac{{({N_y} - 1)}}{2} + 1, - \frac{{({N_y} - 1)}}{2} + 2,...,\frac{{({N_y} - 1)}}{2} - 1$, ${n_z} =  - \frac{{({N_z} - 1)}}{2} + 1, - \frac{{({N_z} - 1)}}{2} + 2,...,\frac{{({N_z} - 1)}}{2} - 1$. Given that the perpendicular direction of the $k^\mathrm{th}$ array is $(\frac{{(k - 1)\pi }}{2},\frac{\pi }{2})$, the response vector of the $k^\mathrm{th}$ array can be thereby written as 
\begin{align}\label{array}
&{{\bf{a}}_k}(\phi ,\theta ) =\notag\\
&\quad\frac{1}{{\sqrt {{N_a}} }}[{e^{j\pi \left[ { - \frac{{({N_y} - 1)}}{2}\sin \left( {\phi  - \frac{{(k - 1)\pi }}{2}} \right)\sin \theta  - \frac{{({N_z} - 1)}}{2}\cos \theta } \right]}},...,\notag\\
&\;\;\;\;\;\;\;\;\;\;\;\;\;\;\;\;\;{e^{j\pi \left[ {{n_y}\sin \left( {\phi  - \frac{{(k - 1)\pi }}{2}} \right)\sin \theta  + {n_z}\cos \theta } \right]}},...,\\
&\qquad \qquad \qquad{e^{j\pi \left[ {\frac{{({N_y} - 1)}}{2}\sin \left( {\phi  - \frac{{(k - 1)\pi }}{2}} \right)\sin \theta  + \frac{{({N_z} - 1)}}{2}\cos \theta } \right]}}{]^T}.\notag
\end{align}

To provide omni-directional communication with adequate array gains, four UPAs are tailored for beamforming in four different space ranges by using the directional antenna elements. As shown in Fig. \ref{mo}(b) and (c), each array is dedicated to transmitting and receiving signals only in the range within $ \pm \frac{\pi }{4}$ to the perpendicular direction of the array, in both azimuth and elevation domains. As such, the transmit/receive range of $k^\mathrm{th}$ array is denoted by
\begin{equation}\label{omi}
{\Omega _k}=\left\{ {({\phi _k},{\theta _k})\left| {\begin{array}{*{20}{c}}
{ - \frac{\pi }{4} + (k - 1)\frac{\pi }{2} \le {\phi _k} \le \frac{\pi }{4} \!+\! (k - 1)\frac{\pi }{2},}\\
{\frac{\pi }{4} \le {\theta _k} \le \frac{{3\pi }}{4}.}
\end{array}} \right.} \right\}.
\end{equation}
Let $s$ denote a transmitted symbol with unit power to the $k^\mathrm{th}$ transmit UPA, the processed received signal from the $m^\mathrm{th}$ receive UPA can be expressed as
\begin{equation}\label{rec}
{y_{k,m}} = \sqrt P {\bf{w}}_m^H {{{\bf{H}}_{k,m}}{{\bf{f}}_k}s}  + {\bf{w}}_m^H{\bf{n}},
\end{equation}
where $P$ represents the transmit power, ${{\bf{H}}_{k,m}}\in {\mathbb{C}^{N_a \times N_a}}$ is the channel matrix between the $k^\mathrm{th}$ transmit UPA and $m^\mathrm{th}$ receive UPA, $s$ is the data symbol,  ${\bf{f}}_k \in {\mathbb{C}^{N_a \times 1}}$ (resp. ${\bf{w}}_m \in {\mathbb{C}^{N_a \times 1}}$) is the normalized beamforming precoder (resp. decoder) at $k^\mathrm{th}$ (resp. $m^\mathrm{th}$) UPA, and ${\bf{n}} \sim \mathcal{CN}({\bf{0}},\sigma^2{\bf{I}})$ is the zero-mean additive Gaussian noise with power $\sigma^2$. Hence, the decoding signal-to-noise ratio (SNR) of $s$ from the $k^\mathrm{th}$ transmit UPA to the $m^\mathrm{th}$ receive UPA is given by 
\begin{equation}\label{rec2}
{\Gamma _{k,m}} = \frac{P}{{{\sigma ^2}}}{\left| {{\bf{w}}_m^H {{{\bf{H}}_{k,m}}{{\bf{f}}_k}} } \right|^2}.
\end{equation}

\subsection{Channel Model}
THz massive MIMO channels generally consist of one LoS path and a few NLoS paths. According to this fact, we adopt the Saleh-Valenzuela
channel model for THz communications. As such, channel ${{\bf{H}}_{k,m}}$ in (\ref{rec}) and (\ref{rec2}) can be further specified as
\begin{subequations}
\begin{align}
&{{\bf{H}}_{k,m}} = {\bf{H}}_{k,m}^{{\rm{LoS}}} + {\bf{H}}_{k,m}^{{\rm{NLoS}}},\\
&{\bf{H}}_{k,m}^{{\rm{LoS}}} = \sqrt{{G_t^k}{G_r^m}{F_k}({\phi _t},{\theta _t}){F_m}({\phi _r},{\theta _r})}\notag\\
&\qquad\qquad\qquad\qquad\qquad\times\alpha _{\rm{L}}{{\bf{a}}_m}({\phi _r},{\theta _r}){{\bf{a}}_k}{({\phi _t},{\theta _t})^H}\label{chan},\\
&{\bf{H}}_{k,m}^{{\rm{NLoS}}}= \;\sum\limits_{l = 1}^{L{\rm{ - }}1} \sqrt{{{G_t^k}{G_r^m}{F_k}(\phi _t^l,\theta _t^l){F_m}(\phi _r^l,\theta _r^l)}}\notag\\
&\qquad\qquad\qquad\qquad\qquad\times\alpha _{\rm{N}}^l {{\bf{a}}_m}(\phi _r^l,\theta _r^l){{\bf{a}}_k}{(\phi _t^l,\theta _t^l)^H},
\end{align}
\end{subequations} 
where ${\bf{H}}_{k,m}^{{\rm{LoS}}}$ and ${\bf{H}}_{k,m}^{{\rm{NLoS}}}$ are the LoS and NLoS components, respectively.  $L$ denotes the number of propagation paths between the transmitter and receiver. ${\alpha}_{{\rm{L}}}$ describes the complex gain of the LoS path and ${\alpha _{{\rm{N}}}^l}$ is the complex gain of the $l^\mathrm{th}$ NLoS path.  ${{\bf{a}}_k}({\phi _t},{\theta _t})$ and ${{\bf{a}}_m}({\phi _r},{\theta _r})$ are the normalized transmit and receive array response vectors, which follow the definition given in (\ref{array}). $ ({\phi _t},{\theta _t})$ and $({\phi _r},{\theta _r})$ (resp. $({\phi _t^l},{\theta _t^l})$ and $({\phi _t^l},{\theta _t^l})$) are the LoS path's (resp. $l^\mathrm{th}$ NLoS path's) azimuth and elevation angles of departure and arrival (AoDs/AoAs), respectively\footnote{We emphasize that different from the convention that the AoDs/AoAs are defined for a single UPA, in this paper, the path angles are defined for the QUPA, as shown in Fig. \ref{mo}.}. ${G_t^k}$ and ${G_r^m}$ are the transmit and receive antenna gains at the $k^\mathrm{th}$ transmit UPA and the $m^\mathrm{th}$ receive UPA, respectively,  which can be written as   
\begin{equation}\label{ff}
G_t^k \;({\mathrm{or}}\;G_r^m)=\frac{{4\pi N_a }}{{\int\limits_{\varphi  = 0}^{2\pi } {\int\limits_{\theta  = 0}^\pi  {F_k(\varphi ,\theta )\sin \theta d\theta d\varphi } } }},
\end{equation}
where ${F_k}({\phi },{\theta })$ is the normalized power radiation pattern of the antenna element at the $k^\mathrm{th}$ UPA.  As each UPA only serves a confined range, higher array gains can be achieved by using the directional antenna element tailored for the certain coverage. To this end, an ideal power radiation pattern of ${F_k}({\phi },{\theta })$ can be expressed as 
\begin{equation}\label{fk}
F_k(\phi ,\theta ) = \left\{ {\begin{array}{*{20}{c}}
{1}, & {(\phi ,\theta )\in {\Omega _k}} 
\\
0 , &{{\rm{otherwise}}}
\end{array} ,} \right.
\end{equation}  
which yields the transmit and receive antenna gains of $G_t^k=G_r^m=4\sqrt{2}N_a$. Moreover, due to the limited angular deflection, i.e., $[ - \frac{\pi }{4},\frac{\pi }{4}]$, the beam squint loss can be effectively reduced in the wideband beamforming. According to \cite{bs1}, the normalized wideband beam gain is the maximum value of the beam patterns' intersection at all frequencies of the signal band, which is given by 
\begin{figure*} 
    \centering 
\includegraphics[width=5.4in]{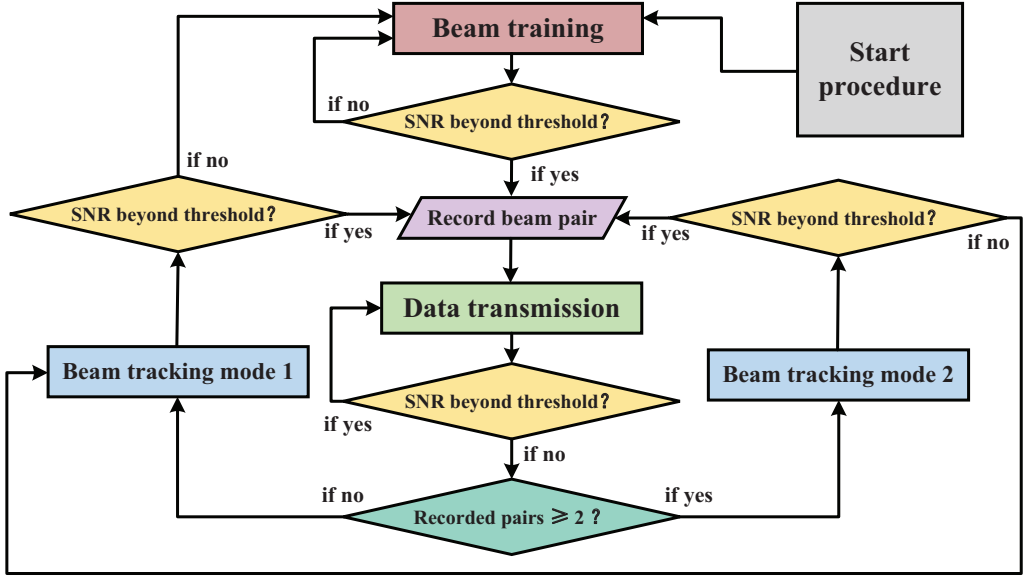}
\caption{The diagram of our proposed framework on unified procedure.}\label{fra}
\vspace{-12pt}
\end{figure*}
\begin{equation}
{A_f}(\varphi ) = \left| {\frac{{\sin (\frac{{\sqrt{N_a}\pi B}}{{4{f_c}}}\sin {\varphi})}}{{\sqrt{N_a}\sin (\frac{{\pi B}}{{4{f_c}}}\sin \varphi )}}} \right|,
\end{equation}
where $B$ is the baseband bandwidth, $\varphi$ is the beam's direction,  and $f_c$ is the carrier frequency. Thus, the maximum beam squint loss can be expressed as $1 - {A_f}({\varphi _{\max }})$, in which $\varphi _{\max }$ is the maximum beam angular deflection. Thus, compared to the conventional UPA, the beam squint loss at QUPA can reduce 
\begin{equation}
L = \frac{{{A_f}(\pi /4) - {A_f}(\pi /2)}}{{1 - {A_f}(\pi /2)}}.
\end{equation}
The reduction $L$ decreases with the increase of $\frac{B}{f_c}$. For example, when $\frac{B}{f_c}=5\%$ (resp. to $20\%$), the QUPA can reduce $49.5\%$ (resp. to $41.4\%$) beam squint loss when $N_a=256$.

\subsection{Problem Statement}
To enable reliable THz communication, the precoder ${\bf{f}}_k$
at the $k^\mathrm{th}$ transmit UPA and the decoder ${\bf{w}}_m$ at the $m^\mathrm{th}$ receive UPA are needed to be optimized under normalized power to maximize the decoding SNR specified in (\ref{rec2}), which is equivalent to solving the following problem
\begin{equation}\label{tran}
\begin{split}
\{ {\bf{w}}_m^{{\rm{opt}}},{\bf{f}}_k^{{\rm{opt}}}\}  = \arg &\max {\left| {{\bf{w}}_m^H{{\bf{H}}_{k,m}}{{\bf{f}}_k}} \right|^2}\\
&{\rm{s}}{\rm{.t}}{\rm{.}}\;\;{\left\| {{{\bf{f}}_k}} \right\|^2} \le 1,\;\;{\left\| {{{\bf{w}}_m}} \right\|^2} \le 1.
\end{split}
\end{equation}
Provided that ${{\bf{H}}_{k,m}}$ is perfectly known at the transmitter and receiver, the optimal precoder ${\bf{f}}_k^{{\rm{opt}}}$ and the decoder ${\bf{w}}_m^{{\rm{opt}}}$ can be easily derived by applying the singular value decomposition on ${{\bf{H}}_{k,m}}$. However, the pilot signals with omnidirectional radiation may not be effectively detected due to the severe path loss in THz channels. In the light of this, we need to find ${\bf{f}}_k^{{\rm{opt}}}$ and ${\bf{w}}_m^{{\rm{opt}}}$ by testing the precoder-decoder pairs (i.e., beam pairs) predefined in a codebook, without any channel state information. This process is referred to as \emph{beam training.}

After obtaining  ${\bf{f}}_k^{{\rm{opt}}}$ and ${\bf{w}}_m^{{\rm{opt}}}$ for ${{\bf{H}}_{i,j}^{{\rm{LoS}}}}$ over a transmission interval $T$, the LoS channel might be changed in the next interval due to the movement (or rotation) of both the transmitter and receiver, i.e., ${\bf{H}}_{k,m}^{{\rm{LoS}}}(T+1) \ne {\bf{H}}_{k,m}^{{\rm{LoS}}}(T)$. Thus, one strategy for maintaining the communication is to re-apply the beam training at the transmission interval $T+1$. However, in practice, the positions of the transmitter and receiver vary gradually, which implies that ${\bf{H}}_{k,m}^{{\rm{LoS}}}(T + 1)$ is closely related to ${\bf{H}}_{k,m}^{{\rm{LoS}}}(T)$. In sight of this, we can find ${\bf{f}}_k^{{\rm{opt}}}(T+1)$ and ${\bf{w}}_m^{{\rm{opt}}}(T+1)$ quickly by testing beam pairs in a reduced codebook based on the prediction of ${\bf{H}}_{i,j}^{{\rm{LoS}}}(T+1)$. This process is referred as \emph{beam tracking.}

In the next sections, we aim to design a unified 3D beam training and tracking procedure for our considered system, with both low computational complexity and time consumption.

\section{Framework on A Unified 3D Beam Training and Tracking Procedure}\label{framew}
In this section, we present a novel communication framework that has dynamic beam training/tracking frequency depending on the real-time communication quality to reduce outages.  Fig. {\ref{fra}} shows the block diagram of the framework on a unified training and tracking procedure. 

This procedure starts with the beam training to find the optimal beam pair to establish reliable communication. Then, with the obtained beam pair, data is transmitted in the subsequent time blocks. When the decoding SNR of the data is lower than a threshold, which indicates that the adopted beam pair is no longer the optimal one, beam tracking mode 1 is applied to find a new beam pair. The beam tracking mode 1 only needs the information of the former recorded beam pair. When a reliable communication link is established, data is transmitted in the subsequent time blocks until the decoding SNR declines below the threshold again. If the number of the recorded beam pairs is greater than $2$, the procedure gives priority to using beam tracking mode 2, which is faster than mode 1, to find a new beam pair. Once beam tracking mode 2 fails to find a reliable beam pair, beam tracking mode 1 will be subsequently applied as a compensation. If both tracking modes are ineffective, the beam training is applied again in the unified procedure. It is worth mentioning that the frequency of applying beam training/tracking is depending on the real-time SNR instead of being fixed. How to set the threshold value will be discussed in Section IV. B. 1).

\section{Beam Training and Tracking}
In this section, we first introduce an exhaustive 3D beam training to show the basic approach of beam alignment. Next, we develop a more efficient GB beam training, including the hierarchical codebook design and the training protocol, to achieve a better performance-complexity trade-off. Finally, we develop a simple yet effective GBH beam tracking that contains two modes of tracking protocol for jointly realizing the fast beam alignment.

\subsection{Beam Training}
\subsubsection{Exhaustive 3D Beam Training}\label{exhs}
Note that the small wavelength and severe path loss significantly limits scattering in THz communication, where the gain of the NLoS paths is much lower than that of the LoS counterpart\cite{ITUR}. Therefore, in this paper, we only consider the LoS component in the beam training. By substituting (\ref{chan})
into (\ref{tran}), the beam training problem is equivalent to
\begin{equation}\label{ori}
\begin{split}
\{ {\bf{w}}_m^{{\rm{opt}}},{\bf{f}}_k^{{\rm{opt}}}\}  = \arg &\max {\left| {{\bf{w}}_m^H\underbrace {{{\bf{a}}_m}({\phi _r},{\theta _r}){{\bf{a}}_k}{{({\phi _t},{\theta _t})}^H}}_{{\rm{can}}\;{\rm{not}}\;{\rm{be}}\;{\rm{obtained}}}{{\bf{f}}_k}} \right|^2}\\
&{\rm{s}}{\rm{.t}}{\rm{.}}\;{{\bf{f}}_k} \in \mathcal{F}_k,{{\bf{w}}_m} \in \mathcal{W}_m.
\end{split}
\end{equation}
Without the codebook constraint, an optimal solution to (\ref{ori}) is given by $\{ {\bf{w}}_m^{{\rm{opt}}} = {{\bf{a}}_m}({\phi _r},{\theta _r}),{\bf{f}}_k^{{\rm{opt}}} = {{\bf{a}}_k}({\phi _t},{\theta _t})\} $. Since the optimal beam pair follow the form of array response vector, a straightforward method to reach a desired solution is to traverse all beam pairs from the codebooks composed of array response vectors with different angles\cite{ywang}. The codebook for the $i^\mathrm{th}$ UPA contains $N^2$ narrow beams (i.e., codewords) with $N$ azimuth angles times $N$ elevation angles uniformly distributed in range $\Omega_i$ (which is specified in (\ref{omi})). This method is also referred to as \emph{exhaustive 3D beam training.} However, when it applies to our considered system with four UPAs, the transmitter and receiver have $4N^2$ narrow beams on each. Thus, there are $16N^4$ beam pairs to be tested in the exhaustive 3D beam training, which is quite time consuming when $N$ is large. Next, we propose a low-complexity yet effective GB beam training including the designs of hierarchical codebook and training protocol.



\begin{figure} 
    \centering 
    \includegraphics[width=3in]{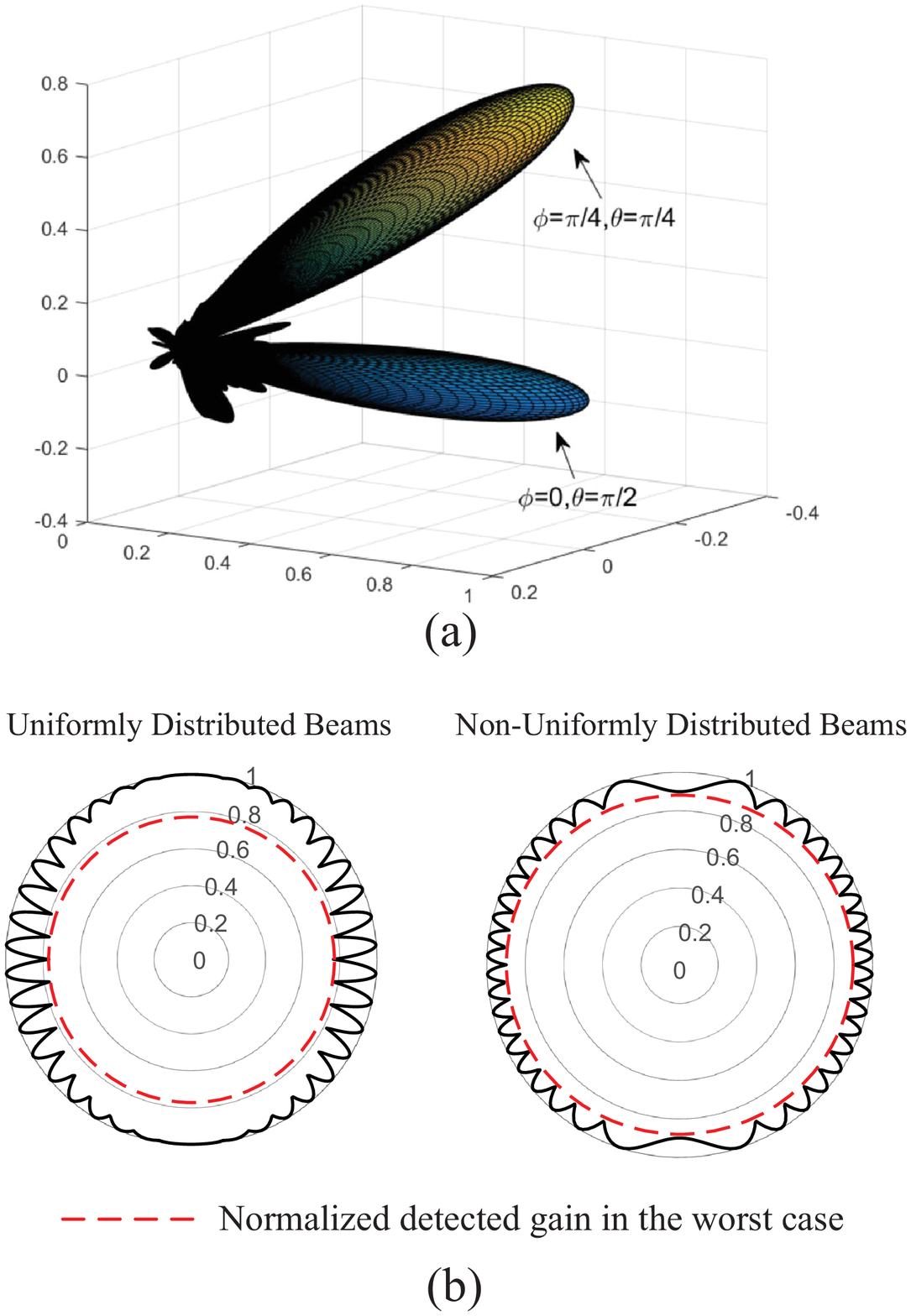} 
   \caption{(a) An example of the beam patterns of two narrow beams. (b) Comparison between uniformly and non-uniformly distributed beams.}\label{2sam}
   \vspace{-12pt}
\end{figure}
\subsubsection{Hierarchical Codebook Design}\label{HCD}
Our proposed 3D hierarchical codebook pre-defines some codewords for narrow beams and wide beams stage-by-stage. The narrow beams act as the solution candidates, which determines the overall training performance. The wide beams are used for identifying the direction of the best narrow beam, which assists to reduce the training complexity. Firstly, we design ${N^2} = {2^S}$ narrow beams that cover $\Omega_k$ in union, where $S$ is the number of stages of our proposed hierarchical codebook. These narrow beams lie in the bottom stage, i.e., stage $S$, and one narrow beam among will be selected as the optimal solution after the beam training. Based on (\ref{ori}), all the narrow beams ought to follow the form of array response vector. As such, the design of the ${N^2}$ narrow beams is reduced to determine their directions, i.e., $\phi$ and $\theta$ in ${{\bf{a}}_k}(\phi ,\theta )$. However, we would like to point out that the direction of these narrow beams should not be uniformly distributed due to the following fact. Assume that an optimal decoder is used in the beam training, based on (\ref{ori}), the received normalized decoding power is given by 
\begin{equation}
\begin{split}
&\mathop {\max }\limits_{{{\bf{f}}_k}} {\left| {{\bf{a}}_k}({\phi _t},{\theta _t})^H{{\bf{f}}_k} \right|^2}\\
&{\rm{s}}.{\rm{t}}.\;{{\bf{f}}_k} \in \mathcal{C}_k^S,
\end{split}
\end{equation}
where $\mathcal{C}_k^S$ represents the $N^2$ narrow beams to be designed in stage $S$  of $\mathcal{C}_k$. Due to the randomness of the wireless channel (random ${\phi _t}$ and ${\theta _t}$), the quality of codewords $\mathcal{C}_k^S$  can be judged by its one-side worst-case performance, i.e.,
 \begin{equation}\label{worst}
\begin{split}
&{\eta _{{\rm{worst}}}}  = \mathop {\min }\limits_{{\phi _t},{\theta _t}} \mathop {\max }\limits_{{{\bf{f}}_k}} {\left|  {{\bf{a}}_k}({\phi _t},{\theta _t})^H{{\bf{f}}_k} \right|^2}\\
&{\rm{s}}.{\rm{t}}.\;{{\bf{f}}_k} \in \mathcal{C}_k^S.
\end{split}
\end{equation}
To analyze the one-side worst-case performance of these narrow beams, we define the normalized narrow beam gain of ${{\bf{a}}_k}(\phi ,\theta )$ in the direction of $({\phi _t},{\theta _t})$ as 

\begin{equation}\label{narrowg}
\begin{split}
A[{{\bf{a}}_k}(\phi ,\theta ),({\phi _t},{\theta _t})] &= \left| {{{\bf{a}}_k}{{(\phi_t ,\theta_t )}^H}{{\bf{a}}_k}({\phi},{\theta})} \right|.
\end{split}
\end{equation}
By plotting the normalized narrow beam gain of ${{\bf{a}}_k}(\phi ,\theta )$ in all directions, we can reach its beam pattern. Interestingly, we notice that the narrow beam is thinner in the boresight direction, while is wider in the directions of coverage edge.  As shown in Fig. \ref{2sam}(a), for the first UPA, the pattern of ${{\bf{a}}_1}(0 , \pi/2)$  is thinner than that of ${{\bf{a}}_1}(\pi/4 , \pi/4)$. In the sight of this, to guarantee a high worst-case performance, the beams in center of $\Omega _k$ should be distributed tightly while that around the edge of $\Omega _k$ should be distributed loosely. For example, Fig. \ref{2sam}(b) shows the radiation patterns of $20$ narrow beams with different colors uniformly or non-uniformly distributed on the $xy$-plane. The non-uniformly distributed beams that are distributed tightly around $\phi=0$ can yield improved worst-case performance. 

Motivated by this, we endeavor to design the directions for narrow beams of $\mathcal{C}_k^S$ to guarantee the highest worst-case performance. As a result, in the bottom stage of our hierarchical codebook, the $N^2$ narrow beams of $\mathcal{C}_k^S$ are given by 
\begin{subequations}\label{15o}
\begin{align}
C_k^S &= \left\{ {{{\bf{a}}_k}\left( {{\phi _n},{\theta _p}} \right)|n,p = 1,2,...,N} \right\},\;\\
{\phi _n} &= \arcsin \left( {\frac{{\sqrt 2 (2n - 1 - N)}}{{2N}}} \right) + \frac{{(k - 1)\pi }}{2},\label{15b}\\
{\theta _p} &= \arccos \left( {\frac{{\sqrt 2 (2p - 1 - N)}}{{2N}}} \right). \label{15c}
\end{align}
\end{subequations}

\begin{proposition}  
If $N^2$ narrow beams (with $N$ azimuth angles times $N$ elevation angles) are adopted to cover $\Omega_k$ in union, the codewords proposed in (\ref{15o}) guarantee the highest worst-case performance (defined in (\ref{worst})), which is given by (normalized by the best-case performance) 
\begin{equation}\label{p1}
{\eta _{{\rm{worst}}}} = \frac{{\sin \left[ {{{\left( {\sqrt 2 {N_z}\pi } \right)} \mathord{\left/
 {\vphantom {{\left( {\sqrt 2 {N_z}\pi } \right)} {4N}}} \right.
 \kern-\nulldelimiterspace} {4N}}} \right]\sin \left[ {{{\left( {\sqrt 2 \beta {N_y}\pi } \right)} \mathord{\left/
 {\vphantom {{\left( {\sqrt 2 \beta {N_y}\pi } \right)} {4N}}} \right.
 \kern-\nulldelimiterspace} {4N}}} \right]}}{N_yN_z{\sin \left[ {{{\left( {\sqrt 2 \pi } \right)} \mathord{\left/
 {\vphantom {{\left( {\sqrt 2 \pi } \right)} {4N}}} \right.
 \kern-\nulldelimiterspace} {4N}}} \right]\sin \left[ {{{\sqrt 2 \beta \pi } \mathord{\left/
 {\vphantom {{\sqrt 2 \beta \pi } {4N}}} \right.
 \kern-\nulldelimiterspace} {4N}}} \right]}},
\end{equation}
where 
\begin{equation}\label{p2}
\beta  = \left\{ {\begin{aligned}
&{1,\quad\qquad\qquad\qquad{\rm{when}}\;N\;{\rm{is}}\;{\rm{odd}}}\\
&{\sin \Big( {\arccos \frac{{\sqrt 2 }}{{2N}}} \Big),\;\;{\rm{when}}\;N\;{\rm{is}}\;{\rm{even}}}
\end{aligned}} \right..
\end{equation}
\end{proposition}

The proof is relegated to Appendix A. We have mentioned that the design of the wide beams in the upper stages is to reduce the training complexity while the training performance is determined by the $N^2$ narrow beams of (\ref{15o}) in the bottom stage.  Thus, Proposition 1 guarantees the normalized worst-case performance of our proposed GB beam training.

\begin{figure}[t]
\centering
\includegraphics[width=3.5in]{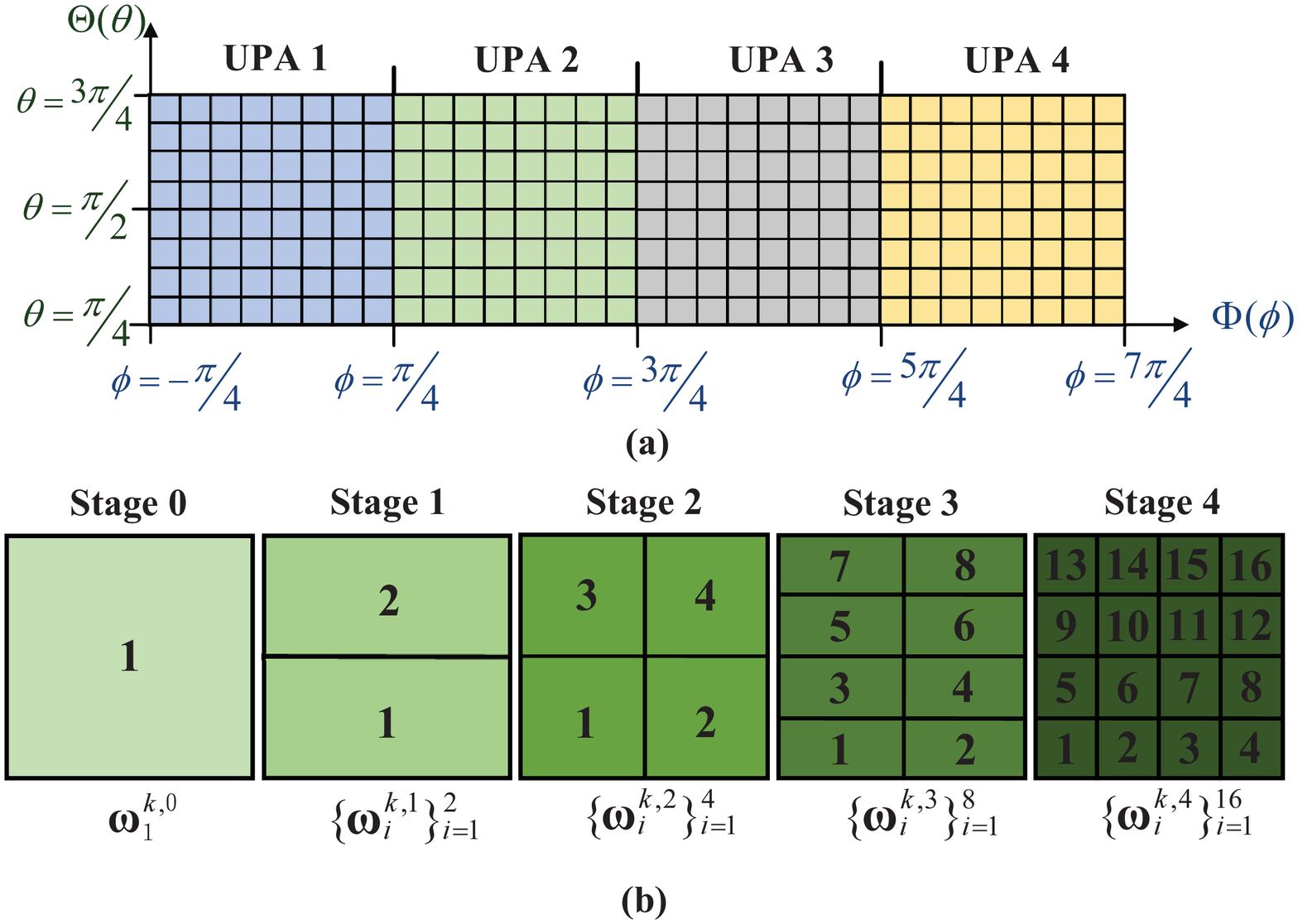}
\caption{(a) An illustration of the range of the narrow beams shown on $\Theta (\theta )$ and $\Phi (\phi )$ on a 3D grid, where $N^2=16$. (b) Beams' distribution and their coverage in different stages on a 3D grid, where $S=4$.}\label{uni}
\vspace{-12pt}
\end{figure}
\begin{figure}[t]
\centering
\includegraphics[width=3.5in]{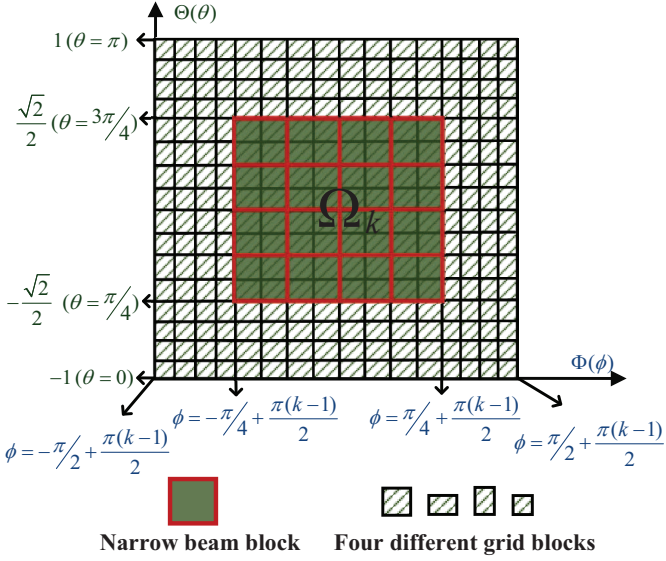}
\caption{A dense grid that represents all directions in front of the $k^ {\mathrm{th}}$ UPA, where $N^2=16$.}\label{gridK}
\vspace{-12pt}
\end{figure}

Next, we introduce our approach to design the wide beams in the upper stages, i.e., stage $0$ to $S-1$. Each wide beam in the stage $s$ covers two beams in the stage $s+1$ while the beam in stage $0$ covers the whole range of $\Omega_k$. As such, we have $2^s$ beams in the stage $s$. The codewords for wide beams are no longer array response vectors and we use $\bm{\omega} _i^{k,s}$ to represent the $i^\mathrm{th}$ codeword in the stage $s$ of the hierarchical codebook $\mathcal{C}_k$.  For ease of illustrating their 3D range, we define two functions as $\Theta (\theta ) =  - \cos \theta$ and $\Phi (\phi ) = \sin (\phi  - \frac{{\pi (k - 1)}}{2}) + \sqrt{2} (k - 1)$. In this way, as shown in Fig. \ref{uni}(a), the range of the narrow beams can be represented by the squares on a 3D grid, where the beam direction is in the center of the square. Based on this representation, Fig. \ref{uni}(b) shows our proposed beams' distribution as well as their coverage in different stages. 
According to the beam index in Fig. \ref{uni}(b) and the beams' distribution in (\ref{15o}), the codewords of the narrow beams can be expressed as
\begin{equation}\label{narrow}
\begin{split}
&\bm{\omega} _i^{k,S} = {{\bf{a}}_k}({\phi _n},{\theta _p}),\;\;i = 1,2,...,{N^2},\\
&n = {\bmod _N}(i),\;\;p = {\rm{ceil}}(i/N),\;\;(\ref{15b}),\;\;(\ref{15c}),
\end{split}
\end{equation}
where $\bmod _N (i)$ returns the remainder after division of $i$ by $N$, and ${\rm{ceil}}(\cdot)$ denotes the ceiling function. To develop the codewords of wide beams for $\mathcal{C}_k$, we have to construct a dense grid that represents all directions in front of the $k^\mathrm{th}$ UPA, i.e.,
\begin{equation}
    \hat\Omega_k \!=\! \Big\{ {(\phi ,\theta )|\phi  \in [ - \frac{\pi }{2} \!+\! \frac{{\pi (k \!-\! 1)}}{2},\frac{\pi }{2} \!+\! \frac{{\pi (k \!-\! 1)}}{2}],\theta  \in [0,\pi ]} \Big\},
\end{equation}
 which is larger than $\Omega_k$. As there are $N\times N$ narrow beams within range $\Omega_k$, as shown in Fig. \ref{gridK}, we construct $2N\times 2N$ grid blocks within this range and total  $4N\times 4N$ grid blocks within $\hat\Omega_k$\footnote{The number of grid blocks can be larger than $4N\times 4N$, which however does not bring noticeable performance gain for the design of wide beams.}.  If each of the rest $4N\!\times\! 4N\!-\!2N\!\times\! 2N$ blocks has the same size of that within  $\Omega_k$, the total coverage is beyond $\hat\Omega_k$. Thus, we set them smaller and uniformly distributed on $\Theta (\theta )$ and $\Phi (\phi )$ to exactly cover  $\hat\Omega_k$. As such, the center directions of the grid blocks for $k$th UPA can be represented as 

\begin{figure*}[t]
\begin{equation}
\left\{ {4N(\beta  \!+\! m) \!+\! \alpha  \!+\! n\left| {\begin{aligned}
&\;{{\rm{when }}\;m = -w,...,-1\;{\rm{and}}\;\delta,...,\delta\!+\!w\!-\!1:\;\;n = 1\!-\!w,...,\nu \! +\! w}\\
&\;\;{{\rm{when }}\;m = 0,1,...,\delta\!-\!1:\;\;n = 1\!-\!w,...,0\;{\rm{and}}\;\nu \! +\! 1,...,\nu\!+\!w\;}
\end{aligned}} \right.} \right\} \tag{28}
\end{equation}
\centering
\includegraphics[width=6.7in]{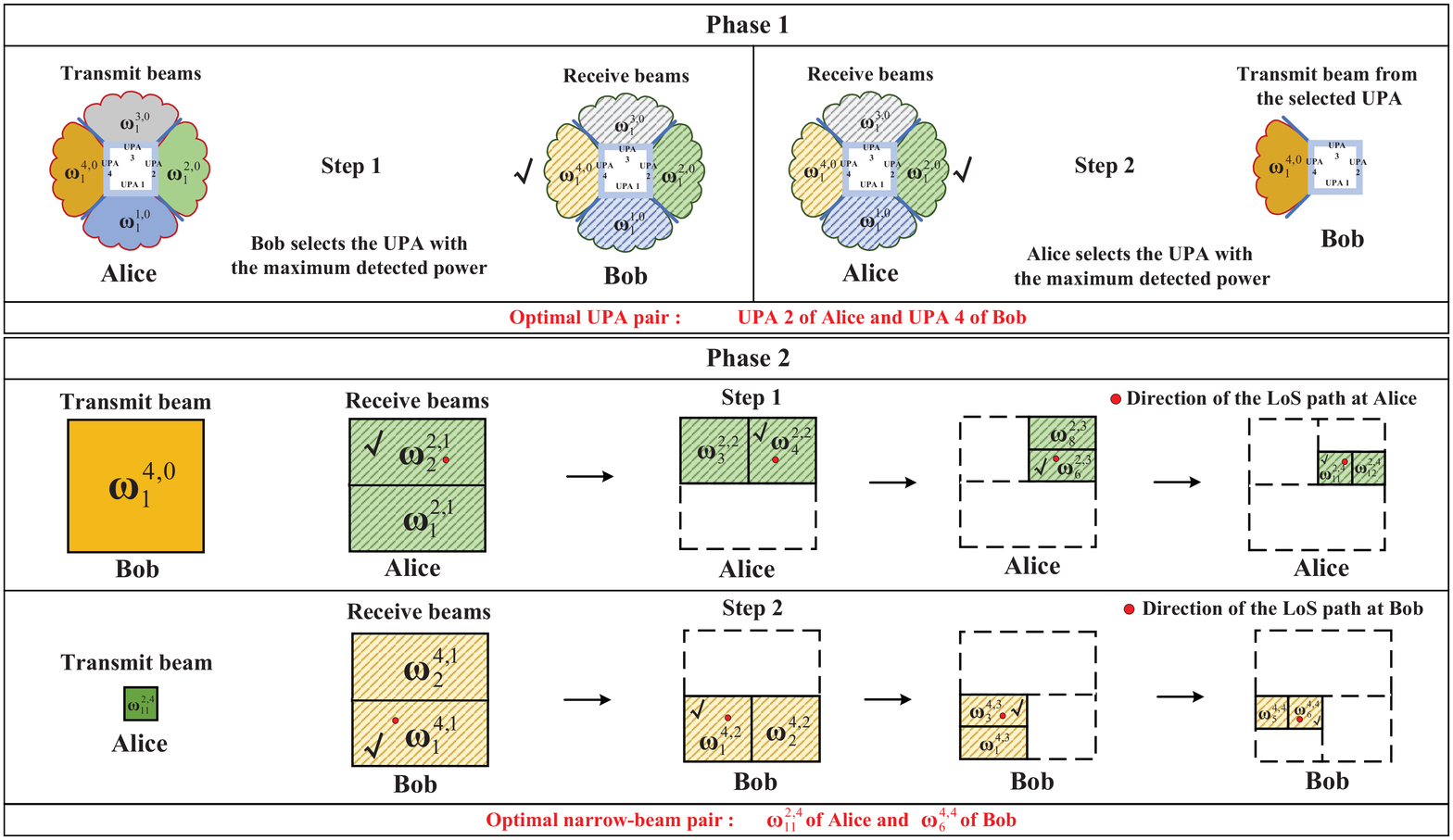}
\caption{A two-phase protocol for the GB beam training, wherein $A^*=2$ and $B^*=4$.}\label{ph1}
\end{figure*}

\begin{small}
\begin{equation*}
\begin{split}
&(\phi _{{\rm{grid}}}^{k,j},\theta _{{\rm{grid}}}^{k,l})\;\;{\rm{with}}\;j= 1,2,...,4N,\;{\rm{and}}\;l= 1,2,...,4N,\\
&\phi _{{\rm{grid}}}^{k,j} \!=\! \left\{ {\begin{split}
&{\arcsin \left( {\frac{{(1 - \sqrt 2 /2)(2j - 1)}}{{2N}} - 1} \right) + \frac{{\pi (k - 1)}}{2},}\\
&{\arcsin \left( {\frac{{\sqrt 2 \left[ {2(j - N) - 1} \right]}}{{4N}} - \frac{{\sqrt 2 }}{2}} \right) + \frac{{\pi (k - 1)}}{2},}\\
&{\arcsin \left( {\frac{{(1 \!-\! \sqrt 2 /2)[2(j\!-\!3N) \!-\! 1]}}{{2N}} \!+\! \frac{{\sqrt 2 }}{2}} \right) \!+\! \frac{{\pi (k \!-\! 1)}}{2},}
\end{split}} \right.
\end{split}
\end{equation*}
\end{small}with piecewise $j = 1,...,N$, $j =N \!+\! 1,...,3N$, and $j =3N \!+\! 1,...,4N$, respectively. 
\begin{small}
\begin{equation*}
\begin{split}
&\theta _{{\rm{grid}}}^{k,l} = \left\{ {\begin{split}
&{\arccos \left( {\frac{{(1 - \sqrt 2 /2)(2l - 1)}}{{2N}} - 1} \right),}\\
&{\arccos \left( {\frac{{\sqrt 2 \left[ {2(l - N) - 1} \right]}}{{4N}} - \frac{{\sqrt 2 }}{2}} \right),}\\
&{\arccos \left( {\frac{{(1 - \sqrt 2 /2)[2(l-3N) - 1]}}{{2N}} + \frac{{\sqrt 2 }}{2}} \right),}
\end{split}} \right.\;\\
\end{split}
\end{equation*}
\end{small}with piecewise $l = 1,...,N$, $l =N \!+\! 1,...,3N,$, and $l =3N \!+\! 1,...,4N$, respectively. According to the proposed beams' distribution as well as their coverage shown in Fig. {\ref{gridK}}, the set of grid directions/blocks covered by  $\bm{\omega} _i^{k,s}$ can be expressed as
\begin{subequations}
\begin{align}
&\Upsilon _i^{k,s} = \left\{ {(\phi _{{\rm{grid}}}^{k,j},\theta _{{\rm{grid}}}^{k,l})\;\left| {j \in \;{J_{s,i}},l \in {L_{s,i}}} \right.} \right\},\\
&{J_{s,i}} \!=\! \left\{ {N \!+\! \nu\cdot {{\bmod }_\mu }(i\!-\!1) \!+\! 1,...,N \!+\! \nu  ({{\bmod }_\mu }(i\!-\!1)\!+\!1)} \right\},\\
&{L_{s,i}} = \left\{ {N + \delta ({\rm{ceil}}\left( {\frac{i}{\mu }} \right) - 1) + 1,...,N + \delta \cdot {\rm{ceil}}\left( {\frac{i}{\mu }} \right)} \right\},\\
&\nu  = {2^{{\rm{ceil}}\left( {\frac{{S - s}}{2}} \right) + 1}},\;\;\delta  = {2^{{\rm{ceil}}\left( {\frac{{S - s - 1}}{2}} \right) + 1}},\;\;\mu  = {2^{{\rm{ceil}}\left( {\frac{{s - 1}}{2}} \right)}},\label{20d}
\end{align}
\end{subequations}  
where $\nu$ represents the number of the grid blocks at the same elevation covered by $\bm{\omega} _i^{k,s}$,  $\delta$ represents the number of the grid blocks at the same azimuth covered by $\bm{\omega} _i^{k,s}$, and $\mu$ represents the number of beams in stage $s$ at the same elevation.
Regarding the wide beams in stage $s$, some prior works \cite{wide1,wide3,thzh,wzhong} expect that $\bm{\omega} _i^{k,s}$ can only achieve beam gain within its coverage $\Upsilon _i^{k,s}$ and cannot achieve the gain in other range, i.e.,
\begin{equation}\label{dc}
{{\bf{a}}_k}{(\phi _{{\rm{grid}}}^{k,j},\theta _{{\rm{grid}}}^{k,l})^H}{\bm{\omega}} _i^{k,s} = \left\{ {\begin{aligned}
&{1,\;\;(\phi _{{\rm{grid}}}^{k,j},\theta _{{\rm{grid}}}^{k,l}) \in \Upsilon _i^{k,s}}\\
&{0,\;\;\;\;\;\;{\rm{otherwise}}\;\;\;\;\;}
\end{aligned}} \right.,
\end{equation}
holds true for all $j=1,2,...,4N$ and $l=1,2,...,4N$. 

However, it is worth mention that the feasible wide beam realized by the beamformer cannot exactly fit (\ref{dc}), and only results in an approximate pattern, which has notable trenches between adjacent ones. This is because the requirement of drastic jump/drop between $0$ and $1$ in (\ref{dc}) may squeeze the resulting beam pattens for minimizing the approximation error. These trenches bring dead zone and impair the overall performance of beam training.  To eliminate them, we modify the criterion given in (\ref{dc}) by adding a buffer zone, which is effective and will be validated in Section \ref{BPC}. The buffer zone ${\rm B}_i^{k,s}$ is the periphery of $\Upsilon _i^{k,s}$ with width of $w$. To write it in the mathematical form, we first set an enlarged zone of $\Upsilon _i^{k,s}$,denoted by $\hat\Upsilon _i^{k,s}$ as
\begin{equation}
\begin{split}
&\hat\Upsilon _i^{k,s} = \left\{ {(\phi _{{\rm{grid}}}^{k,j},\theta _{{\rm{grid}}}^{k,l})\;\left| {j \in \;{\hat J_{s,i}},l \in {\hat L_{s,i}}} \right.} \right\},\\
&{\hat J_{s,i}} = \{N + \nu\cdot {{\bmod }_\mu }(i-1)+1-w,...,\\
&\qquad\qquad\qquad\qquad\quad N + \nu  ({{\bmod }_\mu }(i-1)+1)+w \},\\
&{\hat L_{s,i}} = \left\{ N + \delta ({\rm{ceil}}\left( {\frac{i}{\mu }} \right) - 1)+1-w ,...,\right.\\
&\left.\qquad\qquad\qquad\qquad \quad N + \delta \cdot {\rm{ceil}}\left( {\frac{i}{\mu }} \right)+w \right\},\;(\ref{20d}).\\
\end{split}
\end{equation}
Then, we have ${\rm B}_i^{k,s}=\hat\Upsilon _i^{k,s}-\Upsilon _i^{k,s}$. Thus, in our proposed criterion, we expect that 
\begin{equation}\label{dc2}
{{\bf{a}}_k}{(\phi _{{\rm{grid}}}^{k,j},\theta _{{\rm{grid}}}^{k,l})^H}{\bm{\omega}} _i^{k,s} = \left\{ {\begin{aligned}
&{1,\;\;(\phi _{{\rm{grid}}}^{k,j},\theta _{{\rm{grid}}}^{k,l}) \in \Upsilon _i^{k,s}}\\
&{\chi ,\;\;(\phi _{{\rm{grid}}}^{k,j},\theta _{{\rm{grid}}}^{k,l}) \in {\rm B}_i^{k,s}}\\
&{0,\;\;\;\;\;\;{\rm{otherwise}}\;\;\;\;\;}
\end{aligned}} \right.,
\end{equation}
where $\chi \in (0,1)$ is the expected beam gain in the buffer zone. Define a matrix as
\begin{equation}
\begin{split}
&{\bf{A}} = \left[ {{{\bf{a}}_k}(\phi _{{\rm{grid}}}^{k,1},\theta _{{\rm{grid}}}^{k,1}),...,} \right.\\
&\qquad\qquad\left. {{{\bf{a}}_k}(\phi _{{\rm{grid}}}^{k,4N},\theta _{{\rm{grid}}}^{k,1}),...,{{\bf{a}}_k}(\phi _{{\rm{grid}}}^{k,4N},\theta _{{\rm{grid}}}^{k,4N})} \right].
\end{split}
\end{equation}
Then, we can rewrite (\ref{dc2}) in a more compact form as 
\begin{equation}
{{\bf{A}}^H}[{\bm{\omega}} _1^{k,s},{\bm{\omega}} _2^{k,s},...,{\bm{\omega}} _{{2^s}}^{k,s}] = {\bm{\Xi} _s},
\end{equation}
where ${\bm{\Xi} _s}$ is a $16N^2 \times 2^s$ matrix. The $i^\mathrm{th}$ column of ${\bm{\Xi} _s}$ has an element of 1 in the rows 
\begin{equation}
\left\{ {4N(\beta  + m) + \alpha  + n\left| {m = 0,1,...,\delta  - 1;n = 1,2,...,\nu } \right.} \right\},
\end{equation}
in which $\alpha  = N + \nu \cdot{\bmod _\mu }(i-1)$, $\beta  = N + \delta ({\rm{ceil}}\left( {\frac{i}{\mu }} \right) - 1)$, $\delta$ and $\nu$ are defined in (\ref{20d}), whereas it has an element of $\chi$ in the rows of
(28) in the top of the next page, and has an element of $0$ in other rows. As a result, the $i^\mathrm{th}$ codeword in the stage  $s=0,1,...,S-1$ of $\mathcal{C}_k$ can be obtained as 
\setcounter{equation}{28}
\begin{equation}\label{wide}
{\bm{\omega}} _i^{k,s} = {({\bf{A}}{{\bf{A}}^H})^{ - 1}}{\bf{A}}{\bf{\Xi}} _s(:,i).
\end{equation}

So far, we have obtained all the codewords in the hierarchical codebook $\mathcal{C}_k$. The narrow beams in stage $S$ are given by (\ref{narrow}) and the wide beams in stage $s=0,1,...,S-1$ are given by (\ref{wide}). Next, we propose a low-complexity training protocol for our considered system. For ease of exposition, we call the two nodes as Alice and Bob respectively.
\subsubsection{Beam Training Protocol}
As shown in Fig. {\ref{ph1}}, two phases are developed to achieve different groups of measurements. In Phase 1, we find the optimal UPA pair whose beam range covers the LoS path. Two similar steps are carried to obtain the optimal UPA at Bob and Alice respectively. In step 1, Alice simultaneously uses all UPAs to transmit wide beams via the precoder of ${\bm{\omega}} _1^{k,0}$ for the $k^\mathrm{th}$ UPA. Meanwhile, Bob simultaneously uses all UPAs to receive wide beams via the decoder of ${\bm{\omega}} _1^{m,0}$ by using the $m^\mathrm{th}$ UPA. Then, Bob compares the power of the decoding signals from the four UPAs and selects the one (labeled as $B^*$) with the maximum received signal power. In step 2, Bob only transmits the wide beam by the selected UPA with the precoder of ${\bm{\omega}} _1^{B^*,0}$. Meanwhile, Alice simultaneously uses all UPAs to receive wide beams via the decoder of ${\bm{\omega}} _1^{m,0}$ for the $m^\mathrm{th}$ UPA. Then, Alice finds the UPA (labeled as $A^*$) with the maximum received signal power in the same way. After the two steps, the optimal UPA pair is obtained as the $A^*$th UPA of Alice and the $B^*$th UPA of Bob. In Phase 2, we aim to find the optimal narrow-beam pair between $A^*$th UPA of Alice and the $B^*$th UPA of Bob. Two similar steps are carried to obtain the optimal narrow beam at Bob and Alice respectively.  Step 1 of phase 2 follows step 2 of phase 1, in which Bob transmits a wide beam via the precoder of ${\bm{\omega}} _1^{B^*,0}$ for $B^*$th UPA. Meanwhile, Alice uses the $A^*$th UPA to receive wide beams via testing some codewords in $\mathcal{C}_{A^*}$ from stage $1$ to stage $S$. In each stage, Alice tests two beams and selects the one with larger detected power and in the next stage, Alice tests two beams that are within the range of the selected beam. By recursively repeating this way, Alice can reach a narrow beam (labeled as $a^*$) in the stage $S$. In the step 2, Alice transmits a narrow beam via the precoder of ${\bm{\omega}} _{a^*}^{A^*,S}$ for $A^*$th UPA. Meanwhile, Bob uses the $B^*$th UPA to hierarchically test codewords in $\mathcal{C}_{B^*}$ similarly, and reach a narrow beam (labeled as $b^*$) in the stage $S$. After the two steps, the optimal narrow-beam pair is obtained as ${\bm{\omega}} _{a^*}^{A^*,S}$ of Alice and ${\bm{\omega}} _{b^*}^{B^*,S}$ of Bob.

\subsection{Beam Tracking}\label{trackingpro}
\begin{figure*}[t]
\centering
\includegraphics[width=6.2in]{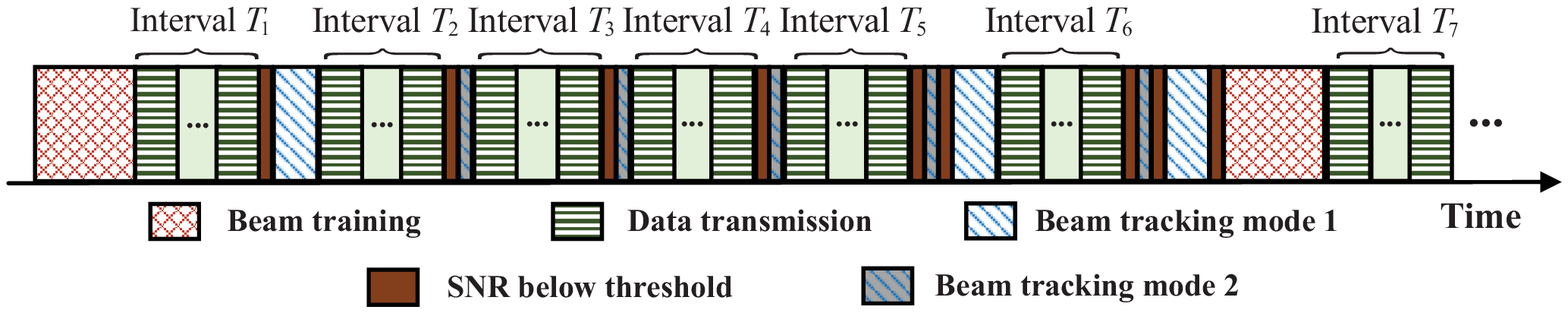}
\caption{An example of operation by our unified training and tracking procedure on the timeline.}\label{epl}
\vspace{10pt}
\centering
\includegraphics[width=6.8in]{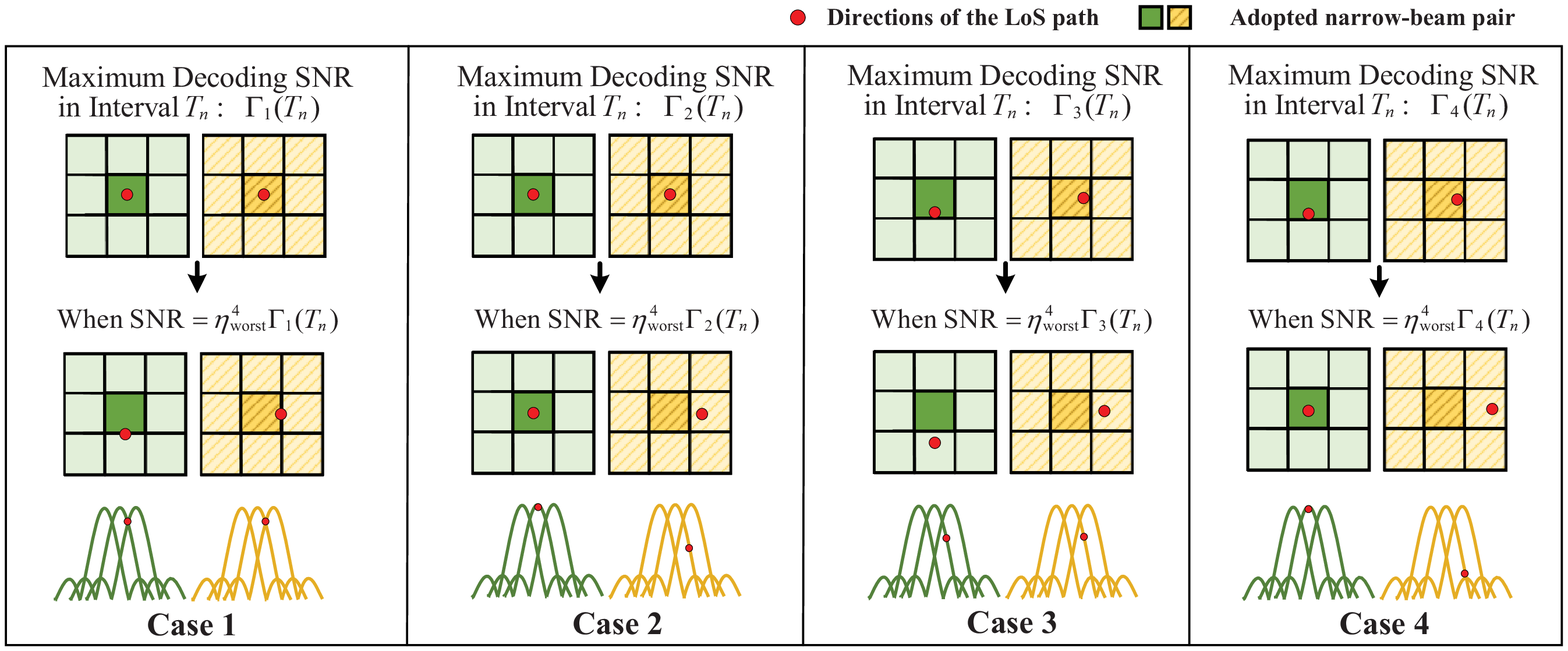}
\caption{The possible path variation between two intervals in four different cases. }\label{dir}
\vspace{-12pt}
\end{figure*}

In this subsection, we propose a low-complexity GBH beam tracking to find the optimal narrow-beam pair in a faster way. It combines two tracking modes with different search times. The first mode needs to search the beams in the vicinity of the former used beam pair, while the second one directly chooses a new beam pair for connection based on the changing trend of the previously used beam pairs. Fig. \ref{epl} shows an example of our unified procedure operation on the timeline. When an aligned beam pair is adopted, we call the period of the subsequent data-transmission time blocks an interval. The decoding SNR at the end of each interval is considered below a threshold and will trigger a new beam tracking. Before developing the beam tracking, we first determine the decoding SNR threshold. 
\subsubsection{Decoding SNR threshold}
With the fixed channel and transit power, we assume that the decoding SNR is merely determined by the beamforming, and thus the SNR will be used to identify the quality of current beam pair. In practice, the SNR may fluctuate occasionally due to the instability of RF devices, same-frequency interference, and etc. In this case, we should consider the effective decoding SNR within a time window, rather than the instant decoding SNR. Denote the optimal narrow-beam pair in the interval $T_n$ by ${{{\bf{\bar w}}}_n}$ and ${{{\bf{\bar f}}}_n}$. Based on (\ref{rec2}), the maximum decoding SNR in the interval $T_n$ can be expressed as 
\begin{equation}
\begin{split}
&\Gamma ({T_n}) = \mathop {\max }\limits_t \frac{P}{{{\sigma ^2}}}{\left| {{\bf{\bar w}}_n^H{\bf{H}}(t){{{\bf{\bar f}}}_n}} \right|^2}\;\\
&\;\;\;\;\;\;\;\;\;\;\;\;{\rm{s}}{\rm{.t}}{\rm{.}}\;\;t \in \;{\rm{Interval }}\;{T_n}\;.
\end{split}
\end{equation}
 Here, we propose a reasonable threshold via the following proposition, proved in Appendix B.
\begin{proposition}
The decoding SNR threshold in interval $T_n$ can be set as $\eta_{{\rm{worst}}}^4\Gamma ({T_n})$, which guarantees that the current beam pair is no longer the optimal one. $\eta _{\rm{worst}}$ is given by (\ref{p1}).
\end{proposition}

Proposition 2 provides a reasonable threshold for practical implementation. This threshold is not fixed but depends on the maximum decoding SNR in each interval. Since $\Gamma ({T_n})$ cannot be determined before the end of the interval, the corresponding threshold may change during the interval $T_n$. Next, we discuss the possible path directions in a new interval. 
\subsubsection{The Possible Path Directions in a New Interval}
\begin{figure*}[t]
\centering
\includegraphics[width=6.8in]{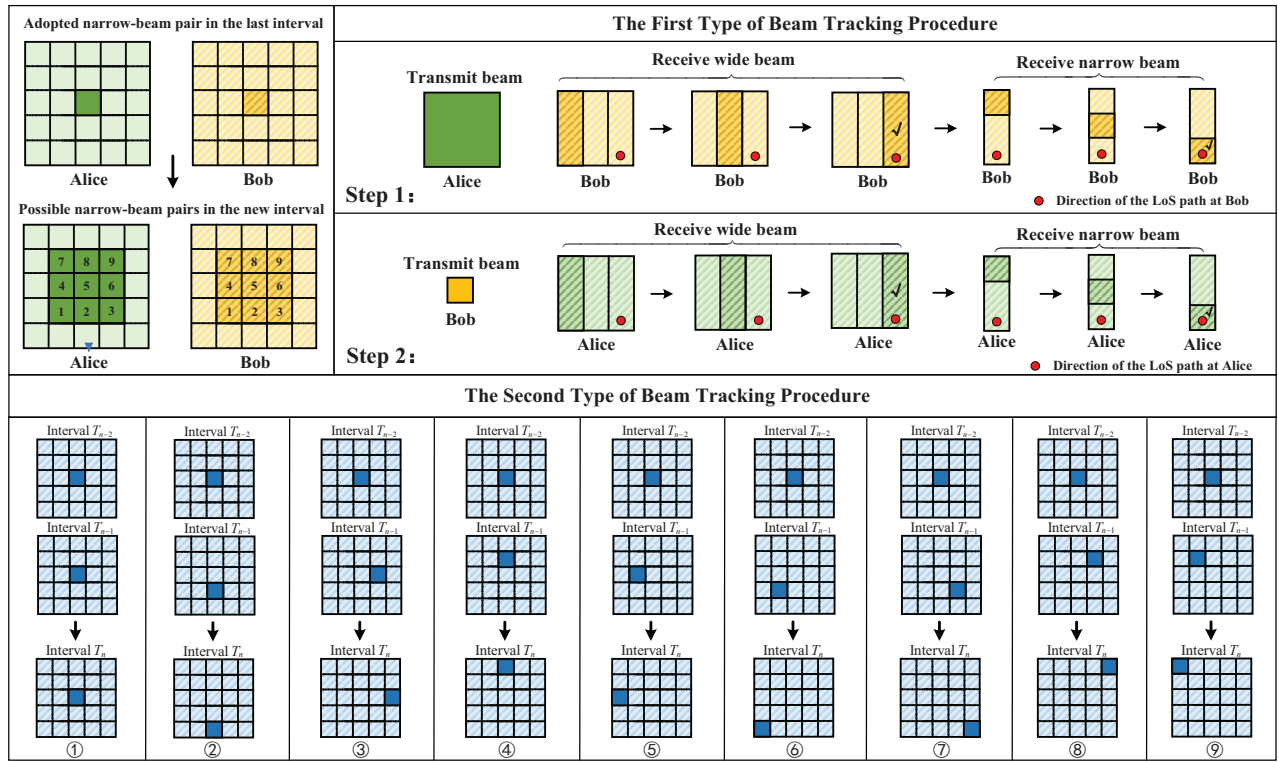}
\caption{A protocol of the two tracking modes for the GBH beam tracking. }\label{tr1}
\vspace{-12pt}
\end{figure*}

When the decoding SNR is less than the threshold $\eta _{{\rm{worst}}}^4\Gamma ({T_n})$, the direction of the LoS path lies outside the range of both ${{{\bf{\bar w}}}_n}$ and ${{{\bf{\bar f}}}_n}$. Fig. \ref{dir} presents four examples of different cases of possible path directions in a new interval.
\begin{itemize}
\item In case 1, the path directions at the maximum decoding SNR in interval $T_n$ are in the center of narrow-beam pair. When the decoding SNR is $\eta _{{\rm{worst}}}^4\Gamma_1 ({T_n})$, both path directions are on the coverage edge. 
\item In case 2, the path directions at the maximum decoding SNR in interval $T_n$ are in the center of narrow-beam pair, i.e., $\Gamma_2 ({T_n})=\Gamma_1 ({T_n})$. When the decoding SNR is $\eta _{{\rm{worst}}}^4\Gamma_2 ({T_n})$, one direction is within the range and the other one is out of the range.
\item In case 3, the path directions at the maximum decoding SNR in interval $T_n$ are not in the center of narrow-beam pair, i.e., $\Gamma_3 ({T_n})<\Gamma_1 ({T_n})$. When the decoding SNR is $\eta _{{\rm{worst}}}^4\Gamma_3 ({T_n})$, both path directions are out of the coverage edge.
\item In case 4, the path directions at the maximum decoding SNR in interval $T_n$ are not in the center of narrow-beam pair, i.e., $\Gamma_4 ({T_n})<\Gamma_1 ({T_n})$. When the decoding SNR is $\eta _{{\rm{worst}}}^4\Gamma_4 ({T_n})$, one direction is within the range and the other one is out of the range.
\end{itemize}
The four cases indicate that in a new interval, the optimal narrow beam on one side must be the original one in the last interval or a neighbor one. As such, there are $9\times9$ candidates in the new interval, one of which is the optimal narrow-beam pair. The optimal solution can be effectively found by our proposed GBH beam tracking, whose protocol of two tracking modes is presented in Fig. {\ref{tr1}}.
\subsubsection{The First Tracking Mode} 
Based on the optimal narrow-beam pair adopted in interval $T_1$, tracking mode 1 aims to find the new optimal one among the $9\times9$ neighboring alternatives via two steps.  In step 1, Alice transmits a wide beam that covers its $9$ narrow-beam candidates. Meanwhile, Bob successively receives $3$ wide beams and selects the one with the largest received signal power, where each wide beam covers $3$ candidates with the same azimuth angle. Then, Bob successively receives $3$ narrow beams within its range and selects the one with the largest received signal power as the optimal narrow beam. In step 2, Bob transmits the obtained narrow beam. In the meanwhile, Alice successively receives beams in the same manner to acquire its optimal narrow beam. After the two steps, Alice and Bob can relocate the optimal narrow-beam pair in interval $T_1$ via only $12$ tests. It is worth mentioning that the codewords of the wide beams used for beam tracking are selected from a dedicated codebook, which can be easily constructed according to the approach proposed in Section \ref{HCD}.  
\subsubsection{The Second Tracking Mode} 
In interval $T_n$ ($n\ge 3$), we could apply tracking mode 2, which is based on the optimal narrow-beam pairs adopted in the last two intervals, i.e., $T_{n-1}$ and $T_{n-2}$. Considering the narrow beam on one side (i.e., Alice side or Bob side), we assume that the transition of the beams between $T_{n}$ and $T_{n-1}$ is the same as that between $T_{n-1}$ and $T_{n-2}$. This assumption has a high probability in practice since the movement or rotation of the transmitter/receiver usually has a strong correlation in a short period. In the second tracking mode, Alice and Bob respectively test their predicted optimal narrow beam pair based on the prediction shown in Fig. {\ref{tr1}}, wherein the predicted one in interval $T_n$ is shown in $9$ different cases.  If the decoding SNR is above the threshold of the former interval, this implies the testing narrow-beam pair is the optimal one. Hence, the GBH tracking is completed directly. If not, Alice and Bob should subsequently apply tracking mode 1, i.e., seek the optimal one among the $9\times9$ candidates, to complete the GBH tracking.

\subsection{Complexity and Applicability Analysis}

\begin{table}[t]
\centering
\caption{Comparison of Complexity and Applicability to THz massive MIMO.} \label{ta_com}
\vspace{0pt}
\hspace{0.45cm}
\begin{tabular}{|c|c|c|}
\hline
Approaches & \tabincell{c}{Applicability} & \tabincell{c}{Search Complexity} \\ \hline
Exhaustive training     & Yes   & $16{N^4} $      \\ \hline
One-sided search\cite{ones}   & No &  $2N^2$\\ \hline
Parallel search\cite{parallel}  & No    & $\left. 16N^4 \middle/ N_{RF} \right.$ \\ \hline
Two-step training\cite{Qsu} &  No  &  $8 N^2$ \\ \hline
MR training \cite{wzhong} &  No  &  $8\log _2{N^2} + 16$ \\ \hline
Proposed training & Yes    &  $4\log _2{N^2} + 2$\\ \hline
Proposed tracking & Yes    &  $12$ or $1$ \\ \hline
\end{tabular}
\vspace{-10pt}
\end{table}
\begin{figure*}[t]
\centering
\includegraphics[width=6.5in]{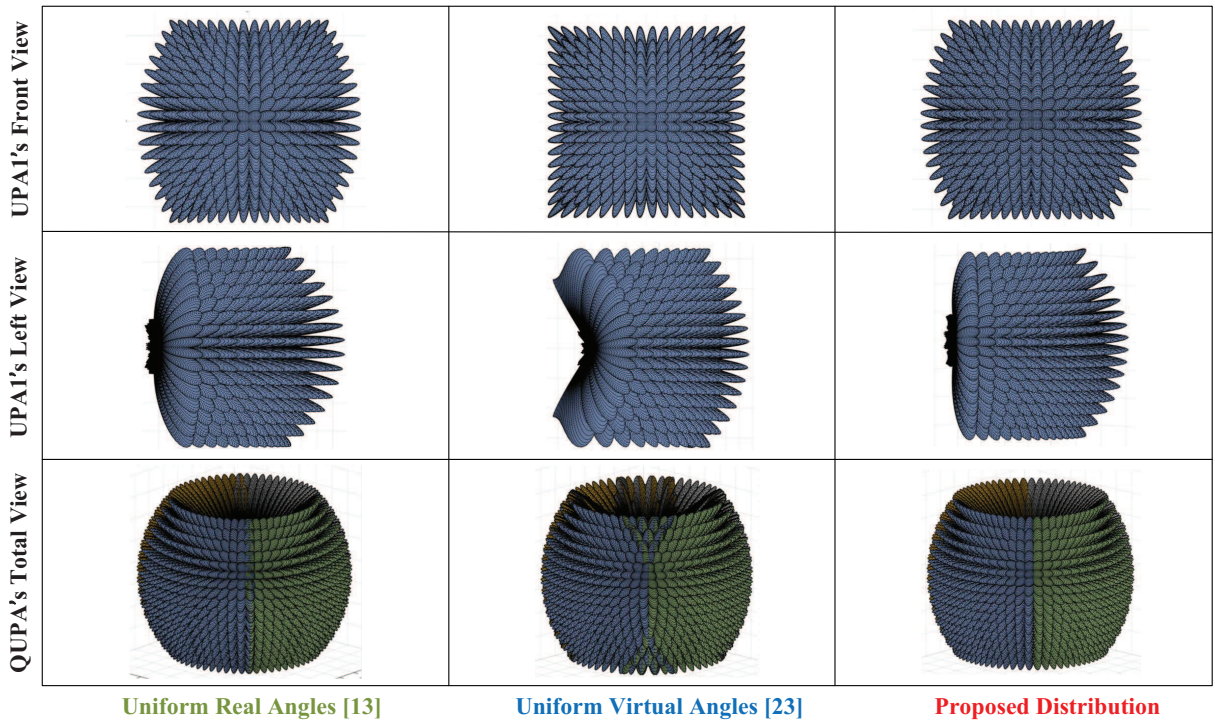}
\caption{Comparison of the proposed narrow beams with the benchmarks, where each UPA has 256 antenna elements.}\label{nacom}
\vspace{-12pt}
\end{figure*}
In this subsection, we compare the search complexity, as well as the applicability to THz massive MIMO, of our proposed beam tracking and training with other 3D training schemes. As mentioned in Section \ref{exhs}, the exhaustive beam training needs to test $16N^4$ beam pairs for our considered systems, which is quite time-consuming when $N$ is large.
To reduce the complexity, IEEE 802.11ad utilizes a one-sided beam search algorithm \cite{ones}, where each user exhaustively searches the narrow beams while the BS transmits the signal in omni-direction, which incurs the complexity of for our considered systems $2N^2$.  The authors in \cite{parallel} proposed a parallel search that uses $N_{RF}$ RF chains at BS to transmit multiple narrow beams simultaneously while all users use an exhaustive training, which incurs the complexity of $\left. 16N^4 \middle/ N_{RF} \right.$ for our considered systems. The authors in \cite{Qsu} proposed a two-step beam training that decomposes the 3D space into $N$ horizontal or vertical sectors (with different elevation/azimuth angles). The two-step beam training has a time complexity of $8N^2$ for our considered systems.  In \cite{wzhong}, a multi-resolution (MR) beam training is proposed by searching the wide-beam pairs first and then the narrow-beam pairs in ${\log _4}(4{N^2})$ stages with $16$ pairs in each stage, which incurs a time complexity of $16{\log _4}(4{N^2})$ for our considered systems. However, omnidirectional beam in \cite{ones}  and simultaneously transmitting multiple beams in \cite{parallel} and \cite{Qsu}  are not practical in THz communication due to the unaffordable transmit power. Moreover, how to realize the wide beams, i.e., the design of the wide-beam codewords, has not been carefully studied in \cite{Qsu} and \cite{wzhong}.

In our proposed beam training, as shown in Fig. \ref{ph1},  there are $4$ tests in Phase 1 and two tests in each stage of Phase 2 that contains $2{\log _2}{N^2}$ stages in two steps. Thus, our proposed beam training has a time complexity of $4{\log _2}{N^2} + 4$. Besides, there is no feedback needed via our scheme while the scheme proposed in \cite{wzhong} needs feedback at every stage. In our proposed beam tracking, as shown in Fig. \ref{tr1}, the first tracking mode requires $12$ beam tests, whereas the second mode requires only one beam test. We summarize the complexity and the applicability of the above approaches in Table \ref{ta_com}.

\section{Numerical Results}\label{nu}
In this section, numerical results are provided to demonstrate the performance of our proposed beam training and tracking. The operating frequency is set to $0.26$ THz and the operating bandwidth is set to $20$ GHz. The communication distance is $100$ m and the noise power spectral density is $-174$ dBm/Hz. Referring to ITU-R P.676-9 \cite{ITUR} and the free space loss formula, the propagation loss is taken as $-124.6$ dB.  According to the first standard at THz band, i.e., IEEE 802.15.3d \cite{vp}, the transmit power is set to $25$ dBm and the antenna gains of $30$ dB are required for mitigating propagation loss. Based on (\ref{ff}) and (\ref{fk}), we use the UPA with $16 \times 16$ elements that incurs the antenna gain of $31.6$ dB.

\subsection{Beam Patterns of the Narrow Beams}\label{BPN}

\begin{figure*}[t]
\centering
\includegraphics[width=6.6in]{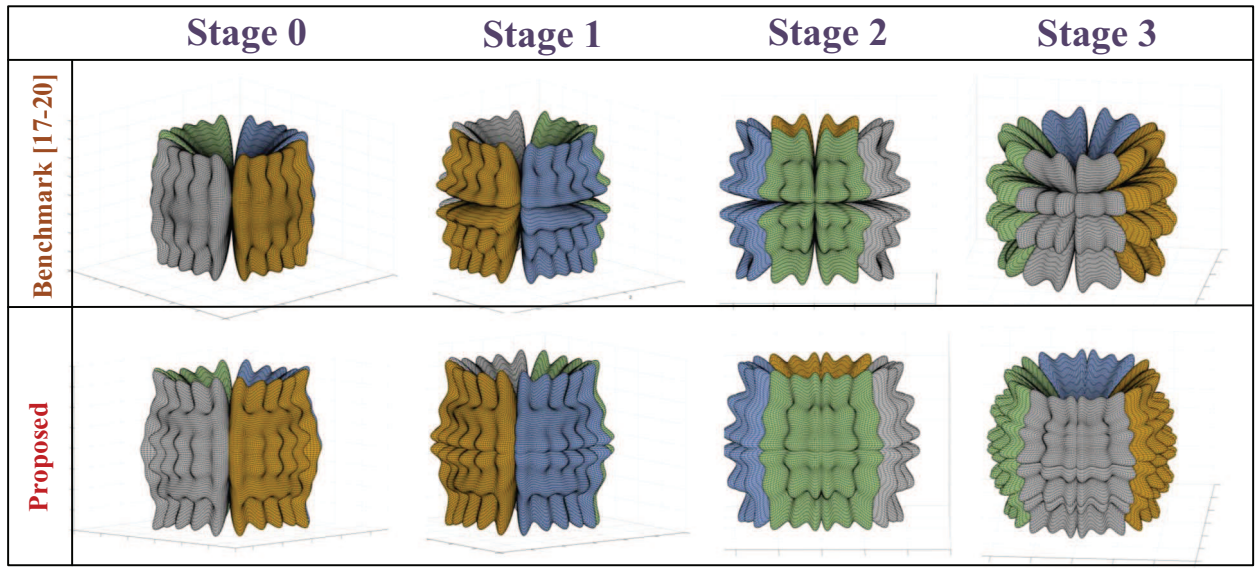}
\caption{Comparison of the proposed wide beams with the benchmarks, where each UPA has 256 antenna elements.}\label{wicom}
\vspace{-12pt}
\end{figure*}
\begin{figure*} 
\centering
\includegraphics[width=7in]{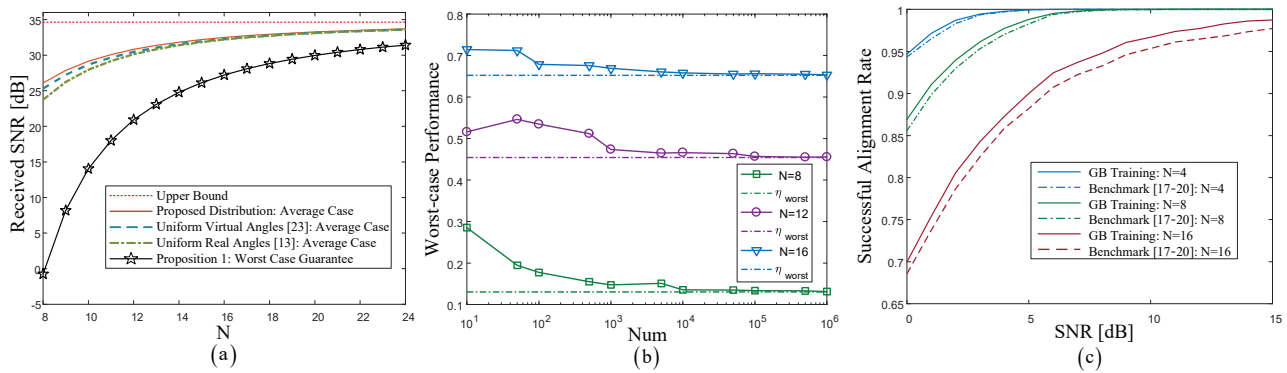}
\caption{(a) Received SNR versus number of narrow beams. (b) Worst-case performance versus test number. (c) Successful alignment rate versus SNR.}\label{ng}
\vspace{-12pt}
\end{figure*}
We first consider the beam patterns of our proposed narrow beams, which are in the bottom stage of the hierarchical codebook. For comparison, we present two benchmarks as follows.
\begin{itemize}
\item {\bf{Uniform Real Angles \cite{ywang}:}} For the $k^\mathrm{th}$ UPA, we extend the codebook in \cite{ywang} to our considered coverage by setting $N$ azimuth angles uniformly distributed in $[ - \frac{\pi }{4} + (k - 1)\frac{\pi }{2},\frac{\pi }{4} + (k - 1)\frac{\pi }{2}]$ and $N$ elevation angles uniformly distributed in $[\frac{\pi }{4},\frac{3\pi }{4}]$.
\item {\bf{Uniform Virtual Angles\cite{wide5}:}} For the first UPA, we consider the simplified array response vector with virtual angles (also known as spatial angles), i.e., 
\begin{equation}
\begin{split}
&{{\bf{a}}_1}(\widetilde \phi ,\widetilde \theta ) = \frac{1}{{\sqrt {{N_a}} }}[1,...,{e^{j\pi ({n_y}\widetilde \phi  + {n_z}\widetilde \theta )}},...,\\
&\qquad\qquad\qquad\qquad\qquad{e^{j\pi [({N_y} - 1)\widetilde \phi  + ({N_z}- 1)\widetilde \theta ]}}]^T,
\end{split}
\end{equation}
where $\widetilde \phi$ and $\widetilde \theta$ are the virtual azimuth and elevation angles within $[-1,1]$. We set $N$ virtual azimuth angles and $N$ virtual elevation angles uniformly distributed $[ - \frac{{\sqrt 2 }}{2},\frac{{\sqrt 2 }}{2}]$. For the $k^\mathrm{th}$ UPA ($k=2,3,4$), we rotate the beam patterns of the first UPA $(k - 1)\frac{\pi }{2}$ in azimuth. As the uniform virtual angles are the optimal distribution for ULA, this benchmark can also be regarded as the Kronecker product scheme extended from the existing 2D codebook\cite{wide5}. 
\end{itemize}

Fig. \ref{nacom} plots the narrow-beam patterns on our proposed codewords in (\ref{15o}) and the benchmarks, where each UPA uses $16 \times 16$ narrow beams to cover its range. For each scheme, three views, i.e., the first UPA's front view (FV), the first UPA's left view (LV), and QUPA's total view (TV), are presented for distinguishing their differences. It can be easily observed from LV that the trenches of the narrow beams with uniform real angles are the deepest, which indicates the lowest worst-case performance. It is interesting to point out that from the FV, although the beams with uniform virtual angles show a figure of a square, the beams at their left and right edges are not vertical. This can be noticed from the LV that the azimuth coverage range will expand when beams are above/below $\frac{\pi}{2}$ of elevation angle. Thus, the total coverage cannot exactly constitute a sphere, which can be seen from the TV that there are some overlaps between adjacent UPAs. Compared to the benchmarks, the beams on our proposed distribution yield the highest worst-case performance, which can be seen from the LV. Moreover, using our proposed distribution, there shows no coverage overlap between different UPAs.

\subsection{Beam Patterns of the Wide Beams}\label{BPC}
Next, we consider the beam patterns of our proposed wide beams $\{\bm{\omega} _i^{k,s}\}_{s=0}^{S-1}$ in (\ref{wide}), which are in stages $2$ to $S$ of the codebook. For comparison, we extend the inverse approach adopted in \cite{wide1,wide3,thzh,wzhong} to the 3D scenario as a benchmark.

Fig. \ref{wicom} plots the wide-beam patterns in stage $0$ to $3$ realized by our proposed approach and the benchmark, respectively, where the adopted hierarchical codebook has $16\times 16$ narrow beams in the bottom stage. It is observed that in stage $0$, the wide beams realized by the two approaches both have notable trenches between adjacent UPAs. However, the trenches of our proposed wide beams are relatively smaller. In stages $1$ to $3$, the wide beams realized by the benchmark all have remarkable trenches even within the coverage range of each UPA. By comparison, there are no trenches within it in the patterns of our proposed wide beams. The beam patterns imply that using our proposed wide beams will have less dead zone during the beam training, and thus are expected to achieve better performance, i.e., higher successful alignment rate.

\subsection{Performance of Beam Training}
\begin{figure*}[t]
\centering
\includegraphics[width=6.8in]{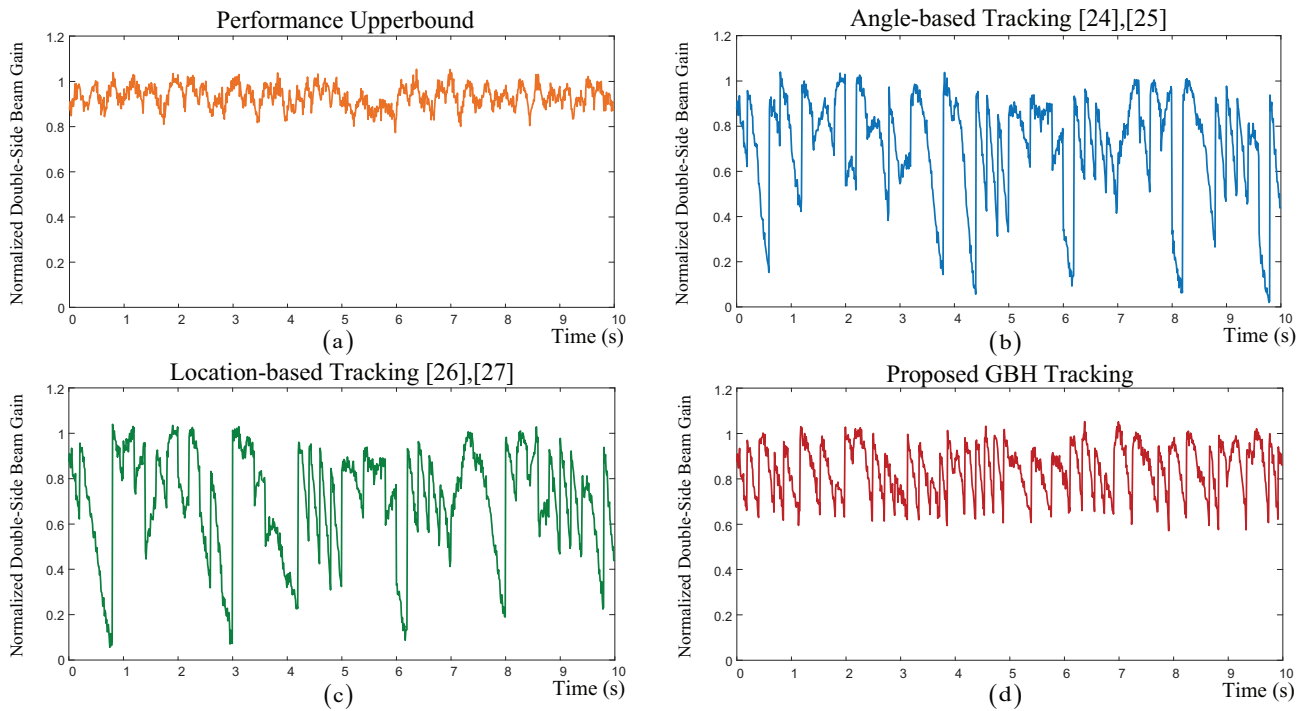}
\caption{The beam tracking performance under four different schemes:
(a) Performance Upperbound.
(b) Angle-based Tracking \cite{tr4,tr5}.
(c) Location-based tracking \cite{tr1,tr3}. 
(d) Proposed GBH tracking.}\label{tkp}
\vspace{-12pt}
\end{figure*}
To validate this point, we evaluate the average/worst-case received SNR after the beam training and the successful alignment rate during the beam training. With a successful alignment, the received SNR is determined by the narrow beams at the bottom stage. Fig. {\ref{ng}}(a) shows the received SNR by using different narrow beams, in which the worst-case guarantee is presented as a baseline. As can be seen, the received SNR of all schemes increases with the number of narrow beams. Our proposed narrow beams outperform the benchmark scheme in \cite{ywang} and \cite{wide5}, where the scheme of using uniform real angles yields the worst performance. With the increase of the number of narrow beams, the gap between the average performance and the worst-case performance decreases.

Next, we validate the theoretical worst-case performance $\eta_{\rm{worst}}$ provided in Proposition 1. To this end, we generate $N_{\rm{test}}$ incoming narrow beams with random AoAs and use the proposed beam training to find the maximum achievable beam gain. Then, we select the lowest one as the worst-case performance in the $N_{\rm{test}}$ tests. Fig. {\ref{ng}}(b) plots the worst-case performance versus the number of tests. It can be observed that for all three different setups, the worst-case performance is gradually approaching $\eta_{\rm{worst}}$ with the increase of the test number. Fig. {\ref{ng}}(c) shows three sets of results (denoted by different colors,
respectively) of the successful alignment rate by using our proposed GB beam training as well as the benchmark scheme\cite{wide1,wide3,thzh,wzhong}. We observe that both schemes can achieve a 100\% successful alignment rate with sufficiently high SNR, and our proposed codebooks can outperform the benchmark codebook in different setups.

\subsection{Performance of Beam Tracking}
We evaluate the performance of our proposed beam tracking. As the tracking has much less search complexity than the training, we consider more narrow beams, i.e., $32 \times 32$ narrow beams, for each UPA to cover their range. To simulate the relative motion between Alice and Bob, we assume that Alice is stationary and staying at $(0,0,0)$ m, whereas Bob is moving from the point of $(100,0,0)$ m. During the motion, we regard Bob as a UAV that randomly changes its moving direction and flies with a maximum speed of $100$ km/h. The total simulation time is $30$ s and each beam test costs $1$ ms. The tracking performance is valued by the changes of the normalized double-side beam gain. 

 Fig. \ref {tkp} shows the performance upperbound and the performance by different schemes. Specifically, we consider the angle-based tracking \cite{tr4,tr5} and location-based tracking \cite{tr1,tr3} as benchmarks. It is worth mentioning that the benchmarks apply the tracking every second periodically, while our proposed GBH beam tracking dynamically applies the procedure based on the SNR threshold given in Proposition 2. The performance upperbound provides an ideal baseline, as we assume that there is a way to accurately obtain the best narrow-beam pair over all the test time. As shown in Fig. \ref {tkp}(a), due to the finite number of beams and background noise, the double-side beam gain cannot maintain to $1$ even in the scheme of upperbound performance. In Fig. \ref {tkp}(b) and (c), the performance of both the angle-based tracking and location-based tracking suffers several inaccurate predictions. This is because Bob's trajectory is connected by multiple segments of linear motion and the above tracking approaches cannot cater for swerves. It can be observed from Fig. \ref {tkp}(d), our proposed GBH beam tracking yields the highest worst-case performance. Assume that the communication outage occurs when the normalized double-side beam gain is below $0.2$. In this case, the angle-based and location-based tracking suffer $6$ and $4$ outages, respectively. By contrast, by combining the first and second tracking modes, no outage occurs via our proposed beam tracking over all the test time.

\section{Conclusions}
We developed a unified 3D beam training and tracking procedure based on a QUPA architecture. To be exact, we first proposed a novel framework to realize the on-demand beam training and tracking with dynamic frequency for THz communication. For realizing beam training, we developed a new hierarchical codebook, in which the narrow beams guarantee the highest worst-case performance and the wide beams have a smaller dead zone. Then, we proposed a low-complexity training protocol to find the optimal narrow-beam pair. As for beam tracking, we developed two tracking modes to jointly realize the beam alignment for mobile transceivers in a fast way. Numerical results plot the 3D beam patterns of the codewords in our proposed codebook, which visually verifies the effectiveness and superiority over benchmark codebooks. Besides, the results show that our proposed GB beam training has advantages on both the beam gain and the successful alignment rate. Our proposed GBH tracking was shown to be able to effectively reduce the outages and maintain adequate beam gain over all the test time. The core of our unified procedure is the proposed framework and training/tracking protocol, based on which the beam codebook can be reconsidered catering for various requirements, e.g., beam coverage\cite{wnrong} and wideband effects\cite{fgao1}. It is also interesting to consider the 3D training and tracking procedure for THz IRS-assisted systems in the future\cite{TIRS1,TIRS2}.

 \begin{appendices}
      \section{Proof of Proposition 1}
\begin{figure}[t]
\centering
\includegraphics[width=2.2in]{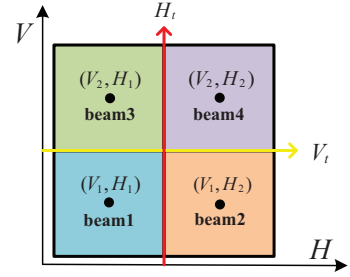}
\caption{The directions of four narrow beams w.r.t. $V$ and $H$.}\label{four}
\vspace{-12pt}
\end{figure}

Note that the directions of beam intersections have relatively lower beam gain. If all the directions of intersections have the same beam gain, the worst-case performance is the highest. Without loss of generality, we discuss the proposed narrow beams for the first UPA, i.e., $\mathcal{C}_1^S$. Note that the normalized narrow beam gain of ${{\bf{a}}_1}(\phi ,\theta )$ defined in (\ref{narrowg}) can be further expressed as 
\begin{equation}
\begin{split}
&A[{{\bf{a}}_1}(\phi ,\theta ),({\phi _t},{\theta _t})]\\
&=\Bigg| \frac{1}{{{N_a}}}\sum\limits_{{n_z} = 0}^{{N_z} - 1} \sum\limits_{{n_y} = 0}^{{N_y} - 1} {e^{j\pi \left[ {\left( {{n_z} - \frac{{({N_z} - 1)}}{2}} \right)\left( {\cos \theta  - \cos {\theta _t}} \right)} \right]}}\\
&\qquad\qquad\qquad\times{e^{j\pi \left[ {\left( {{n_y} - \frac{{({N_y} - 1)}}{2}} \right)\left( {\sin \theta \sin \phi  - \sin {\theta _t}\sin {\phi _t}} \right)} \right]}}   \Bigg|\\
& = \frac{1}{{{N_a}}}\Bigg| \left[ {\sum\limits_{{n_z} = 0}^{{N_z} - 1} {{e^{j\pi \left[ {\left( {{n_z} - \frac{{({N_z} - 1)}}{2}} \right)\left( {\cos \theta  - \cos {\theta _t}} \right)} \right]}}} } \right]\\
&\qquad\qquad\times\left[ {\sum\limits_{{n_y} = 0}^{{N_y} - 1} {{e^{j\pi \left[ {\left( {{n_y} - \frac{{({N_y} - 1)}}{2}} \right)\left( {\sin \theta \sin \phi  - \sin {\theta _t}\sin {\phi _t}} \right)} \right]}}} } \right] \Bigg|\\
& = \frac{1}{{{N_a}}}\Bigg| {e^{ - j\frac{{\pi ({N_z} - 1){m_1}}}{2}}}\frac{{\left( {1 - {e^{j\pi {N_z}{m_1}}}} \right)}}{{1 - {e^{j\pi {m_1}}}}}\\
&\qquad\qquad\qquad\qquad\qquad\quad\times{e^{ - j\frac{{\pi ({N_y} - 1){m_2}}}{2}}}\frac{{\left( {1 - {e^{j\pi {N_y}{m_2}}}} \right)}}{{1 - {e^{j\pi {m_2}}}}} \Bigg|\\
& = \frac{1}{{{N_a}}}\left| {\frac{{\left( {{e^{j\frac{{\pi {N_z}{m_1}}}{2}}} - {e^{ - j\frac{{\pi {N_z}{m_1}}}{2}}}} \right)}}{{\left( {{e^{j\frac{{\pi {m_1}}}{2}}} - {e^{ - j\frac{{\pi {m_1}}}{2}}}} \right)}}\frac{{\left( {{e^{j\frac{{\pi {N_y}{m_2}}}{2}}} - {e^{ - j\frac{{\pi {N_y}{m_2}}}{2}}}} \right)}}{{\left( {{e^{j\frac{{\pi {m_2}}}{2}}} - {e^{ - j\frac{{\pi {m_2}}}{2}}}} \right)}}} \right|\\
& = \frac{1}{{{N_a}}}\left| {\frac{{\sin \left[ {({N_z}\pi {m_1})/2} \right]}}{{\sin \left[ {(\pi {m_1})/2} \right]}}} \right| \cdot \left| {\frac{{\sin \left[ {({N_y}\pi {m_2})/2} \right]}}{{\sin \left[ {(\pi {m_2})/2} \right]}}} \right|,
\end{split}
\end{equation}
where ${m_1} = \cos \theta  - \cos {\theta _t}$ and ${m_2} = \sin \theta \sin \phi  - \sin {\theta _t}\sin {\phi _t}$. Define a two-dimension transformation as
\begin{equation}
\left\{ {\begin{aligned}
&{V\left( \theta  \right) = \cos \theta, \;}\\
&{H(\theta ,\phi ) = \sin \theta \sin \phi, }
\end{aligned}} \right.
\end{equation}
The narrow beam response vector can be expressed as a new vector function that depends on $V$ and $H$, i.e., ${{\bf{a}}_1}(\phi ,\theta ) \Rightarrow {{{\bf{\hat a}}}_1}(V,H)$. As such, the normalized narrow beam gain of ${{\bf{a}}_1}(\phi ,\theta )$ in the direction of $({\phi _t},{\theta _t})$ can be rewritten as that of ${{{\bf{\hat a}}}_1}(V,H)$ in the direction of $({V_t},{H_t})$, i.e.,
\begin{equation}
\begin{split}
A[{{\bf{a}}_1}(\phi ,\theta ),({\phi _t},{\theta _t})] &= A[{{{\bf{\hat a}}}_1}(V,H),({V_t},{H_t})] \\
&= \frac{1}{{{N_a}}}{f_z}(V - {V_t}){f_y}(H - {H_t}),
\end{split}
\end{equation}
where 
\begin{equation}
{f_z}(x) = \left| {\frac{{\sin \left[ {{{\left( {{N_z}\pi x} \right)} \mathord{\left/
 {\vphantom {{\left( {{N_z}\pi x} \right)} 2}} \right.
 \kern-\nulldelimiterspace} 2}} \right]}}{{\sin \left[ {{{\pi x} \mathord{\left/
 {\vphantom {{\pi x} 2}} \right.
 \kern-\nulldelimiterspace} 2}} \right]}}} \right|,{f_y}(x) = \left| {\frac{{\sin \left[ {{{\left( {{N_y}\pi x} \right)} \mathord{\left/
 {\vphantom {{\left( {{N_y}\pi x} \right)} 2}} \right.
 \kern-\nulldelimiterspace} 2}} \right]}}{{\sin \left[ {{{\pi x} \mathord{\left/
 {\vphantom {{\pi x} 2}} \right.
 \kern-\nulldelimiterspace} 2}} \right]}}} \right|.
\end{equation}
Fig. {\ref{four}} illustrates four beams with codewords ${{{\bf{\hat a}}}_k}(V_1,H_1)$, ${{{\bf{\hat a}}}_k}(V_1,H_2)$, ${{{\bf{\hat a}}}_k}(V_2,H_1)$, and ${{{\bf{\hat a}}}_k}(V_2,H_2)$. Assuming that $V_t$ is the direction of intersections between beam 1 and beam 3 (or between beam 2 and beam 4), the two beams should yield the same beam gain on the direction of $V_t$, i.e.,
\begin{subequations}
\begin{align}
& A[{{{\bf{\hat a}}}_k}({V_1},H),({V_t},H)] = A[{{{\bf{\hat a}}}_k}({V_2},H),({V_t},H)]\\
 &\Rightarrow {f_z}({V_1} - {V_t}) = {f_z}({V_2} - {V_t}) \Rightarrow \left| {{V_1} - {V_t}} \right| = \left| {{V_2} - {V_t}} \right|\notag\\
 &\Rightarrow \;{V_t} = \frac{{{V_1} + {V_2}}}{2}.\label{vt}
\end{align}
\end{subequations}
Based on (\ref{vt}), the normalized narrow beam gain on the direction of the intersection between ${{{\bf{\hat a}}}_k}({V_p},H)$ and ${{{\bf{\hat a}}}_k}({V_{p+1}},H)$ can be written as

\begin{align}
A\Big[{{{\bf{\hat a}}}_k}({V_p},H),(\frac{{{V_p} + {V_{p + 1}}}}{2},H)\Big]
= \frac{1}{{{N_a}}}{f_z}(\frac{{{V_p} - {V_{p + 1}}}}{2}){f_y}(0).\label{ann}
\end{align}
To satisfy that all the directions of the intersections have the same beam gain, based on (\ref{ann}), we have that ${f_z}(\frac{{{V_p} - {V_{p + 1}}}}{2})$ is the the same for all $n$, which is equivalent to that $V_{p + 1}-V_p$ is the same for all $n$. According to the range of $V$, we have
\begin{equation}
\theta  \in [\frac{\pi }{4},\frac{{3\pi }}{4}] \Rightarrow V(\theta ) \in [ - \frac{{\sqrt 2 }}{2},\frac{{\sqrt 2 }}{2}].
\end{equation}
Since there are $N$ beams on the elevation, we set $\{V_p\}_{p=1}^N$ satisfying equal $V_{p + 1}-V_p$ within $[ - \frac{{\sqrt 2 }}{2},\frac{{\sqrt 2 }}{2}]$ and regard $\pm \frac{{\sqrt 2 }}{2}$ as the directions of intersections. Consequently, we obtain 
\begin{equation}\label{vv}
{V_p} =  - \frac{{\sqrt 2 }}{2} + \frac{{\sqrt 2 (2p - 1)}}{{2N}},p = 1,2,...,N.
\end{equation}
Similarly, the normalized narrow beam gain on the direction of the intersection between ${{{\bf{\hat a}}}_k}(V,H_n)$ and ${{{\bf{\hat a}}}_k}({V},H_{n+1})$ is determined by $H_{n+1}-H_n$. According to the range of $H$ with fixed $V(\theta)$, we have,
\begin{equation}
\phi  \in [ - \frac{\pi }{4},\frac{\pi }{4}] \Rightarrow H(\theta ,\phi ) \in [ - \frac{{\sqrt 2 }}{2}\sin \theta ,\frac{{\sqrt 2 }}{2}\sin \theta ].
\end{equation}
Since there are $N$ beams on the azimuth, we set $\{H_n\}_{n=1}^N$ satisfying equal $H_{n + 1}-H_n$ within $[ - \frac{{\sqrt 2 }}{2}\sin \theta ,\frac{{\sqrt 2 }}{2}\sin \theta ]$ and regard $\pm \frac{{\sqrt 2 }\sin \theta}{2}$ as the directions of intersections. Consequently, we obtain 
\begin{equation}\label{hh}
{H_n}(\theta) =  - \frac{{\sqrt 2 }}{2} + \frac{{\sqrt 2 (2n - 1)\sin \theta }}{{2N}},n = 1,2,...,N.
\end{equation}
By leveraging the following inverse transformation,
\begin{equation}
\left\{ {\begin{aligned}
&{{\theta _p} = \arccos {V_p}},\qquad p=1,2,...,N\\
&{{\phi _n} = \arcsin \frac{{{H_n}}}{{\sin \theta  }}},\quad \;n=1,2,...,N
\end{aligned}}  \right. ,  
\end{equation}
we get the $N \times N$ narrow beams shown in (\ref{15o}). By this design, for any $N$ beams distributed with the same elevation angle $\theta$, all the directions of intersections have the same gain, denoted by $\eta_{\theta}$. However, it is interesting to point out that the same gain of intersections can be only guaranteed in terms of the 2D azimuth plane since $\eta_{\theta}$ changes with $\theta$.

The corresponding worst-case performance is the normalized narrow beam gain on the direction of intersection between ${{{\bf{\hat a}}}_k}({V_n},H_n(\theta _n))$ and ${{{\bf{\hat a}}}_k}({V_{n+1}},H_{n+1}(\theta _{n+1}))$, i.e.,
\begin{equation}\label{2277}
\begin{split}
&A\Big[{{{\bf{\hat a}}}_k}({V_n},{H_n}),(\frac{{{V_n} + {V_{n + 1}}}}{2},\frac{{{H_n} + {H_{n + 1}}}}{2})\Big]\\
&\qquad\qquad\quad= \frac{1}{{{N_a}}}{f_z}(\frac{{{V_n} - {V_{n + 1}}}}{2}){f_y}(\frac{{{H_n} - {H_{n + 1}}}}{2}).
\end{split}
\end{equation}
Define a function $G(v,h)=\frac{1}{{{N_a}}}{f_z}(v){f_y}(h)$. By substituting (\ref{vv}) and (\ref{hh}) into (\ref{2277}), we have  
\begin{equation}\label{worstt}
\begin{split}
&A\Big[{{{\bf{\hat a}}}_k}({V_n},{H_n}),(\frac{{{V_n} + {V_{n + 1}}}}{2},\frac{{{H_n} + {H_{n + 1}}}}{2})\Big] \\
&\qquad\qquad\qquad\qquad\qquad\qquad= G\Big( {\frac{{\sqrt 2 }}{{2N}},\frac{{\sqrt 2 \sin \theta }}{{2N}}} \Big).
\end{split}
\end{equation}
Equation (\ref{worstt}) indicates that if $N$ beams are distributed with the same elevation angle $\theta$, the worst-case performance of these beams decreases when $\theta$ close to $\pi/2$. Based on (\ref{vv}) and (\ref{2277}), the elevation angles $\{\theta_p\}_{p=1}^N$ are given by (\ref{15c}). Thus, when $N$ is odd, the closet $\theta_p$ is with $p=\frac{N+1}{2}$. When $N$ is even, the closet $\theta_p$ is with $p=\frac{N}{2}$ or $p=\frac{N}{2}+1$.  Thereby, we obtain the worst-case performance of the all $N^2$ beams as given in (\ref{p1}) and (\ref{p2}).
   \section{Proof of Proposition 2}
Assume that the directions of the LoS path at the maximum decoding SNR in the interval $T_n$ are exactly in the center of the range of the optimal narrow-beam pair. In this case, the decoding SNR is the highest, i.e.,
\begin{equation}\label{e27}
{\Gamma ^*}({T_n}){\rm{ = }}\frac{P}{{{\sigma ^2}}}{\left| {{\bf{\bar w}}_n^H{{\bf{H}}^{\rm{*}}}{{{\bf{\bar f}}}_n}} \right|^2} \ge \Gamma ({T_n}){\rm{ = }}\frac{P}{{{\sigma ^2}}}{\left| {{\bf{\bar w}}_n^H{\bf{H}}{{{\bf{\bar f}}}_n}} \right|^2}.
\end{equation}
When the directions of the LoS path are on the coverage edges of both ${{{\bf{\bar w}}}_n}$ and ${{{\bf{\bar f}}}_n}$, the decoding SNR holds that
\begin{equation}\label{e28}
\Gamma {\rm{ = }}\frac{P}{{{\sigma ^2}}}{\left| {({\eta _{{\rm{worst}}}}{\bf{\bar w}}_n^H){{\bf{H}}^{\rm{*}}}({\eta _{{\rm{worst}}}}{{{\bf{\bar f}}}_n})} \right|^2}{\rm{ = }}\eta _{{\rm{worst}}}^4{\Gamma ^{\rm{*}}}({T_n}).
\end{equation}
Thus, if ${{{\bf{\bar w}}}_n}$ and ${{{\bf{\bar f}}}_n}$ are not the optimal narrow-beam pair, the decoding SNR should be less than $\eta _{{\rm{worst}}}^4{\Gamma ^{\rm{*}}}({T_n})$. However, ${\Gamma ^{\rm{*}}}({T_n})$ is unavailable in practice. As a result, we choose a lower bound, i.e., $\eta _{{\rm{worst}}}^4\Gamma ({T_n})$, as an alternative based on the inequality in ({\ref{e27}}). 

\end{appendices}

\begin{IEEEbiography}[{\includegraphics[width=1in,clip]{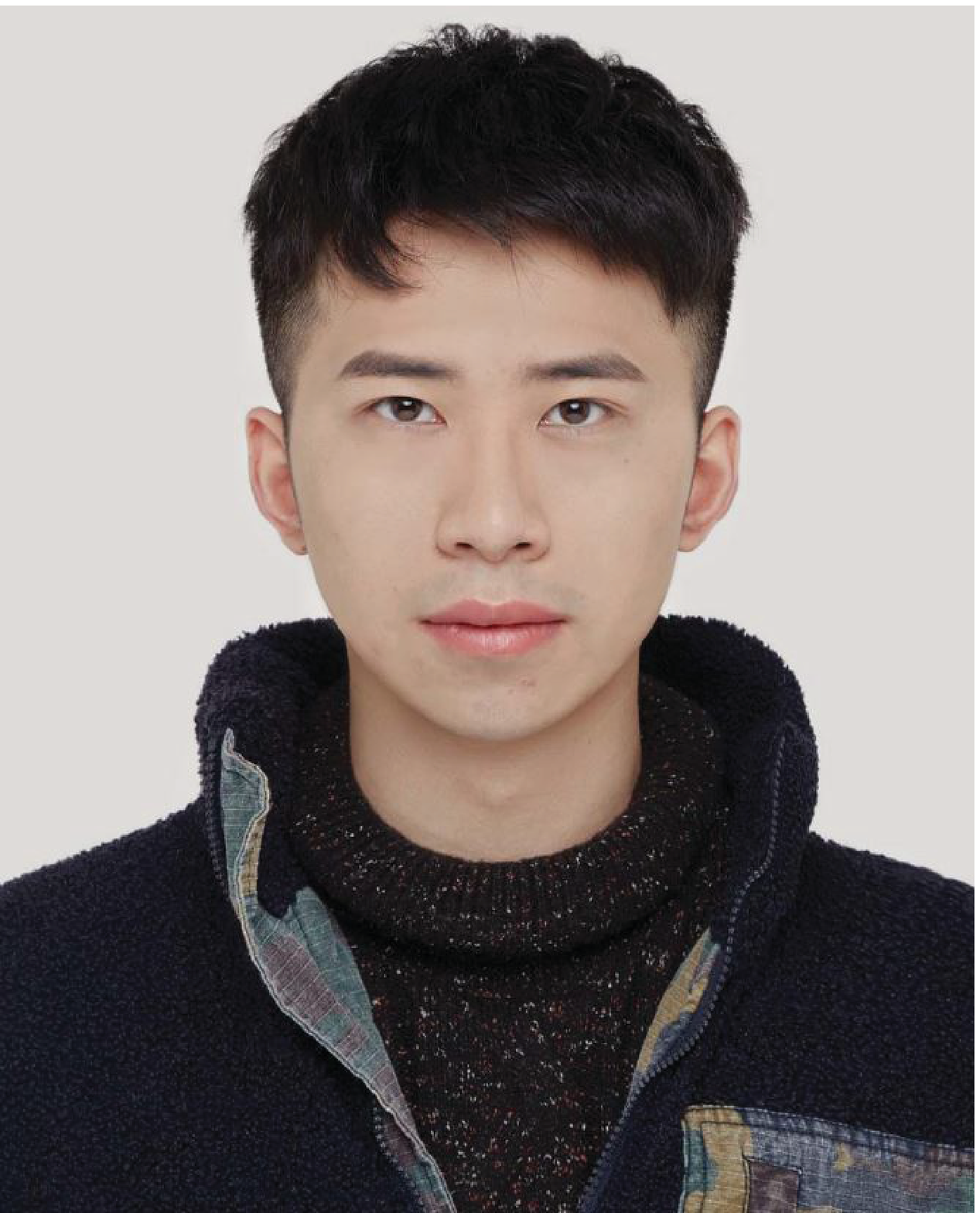}}]{Boyu Ning} received the B.E. degree in Communication Engineering, along with the Certification of the Talent Program in Yingcai Honors College, from the University of Electronic Science and Technology of China (UESTC), Chengdu, China, in 2018. He won the Most Comprehensive Scientific Research Award of the Oxford Study Programme from the Oxford University as a visiting student, in 2018. He was a recipient of the ``Tang Lixin'' Scholarship, in 2019. He is currently pursuing the Ph.D. degree with the National Key Laboratory of Science and Technology on Communications, UESTC. His research interests include Terahertz communication, intelligent reflecting surface, massive MIMO, physical-layer security, and convex optimization.
\end{IEEEbiography}

\begin{IEEEbiography}
[{\includegraphics[width=1in,clip]{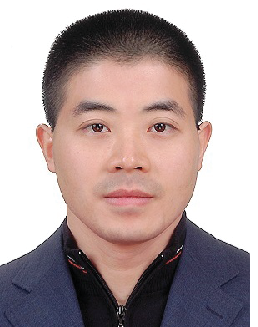}}]{Zhi Chen} received B. Eng, M. Eng., and Ph.D. degree in Electrical Engineering from University of Electronic Science and Technology of China (UESTC), in 1997, 2000, 2006, respectively. On April 2006, he joined the National Key Lab of Science and Technology on Communications (NCL), UESTC, and worked as a professor in this lab from August 2013. He was a visiting scholar at University of California, Riverside during 2010-2011. He is also the deputy director of Key Laboratory of Terahertz Technology, Ministry of Education. His current research interests include Terahertz communication, 5G mobile communications and tactile internet. 
\end{IEEEbiography}

\begin{IEEEbiography}
[{\includegraphics[width=1in,clip]{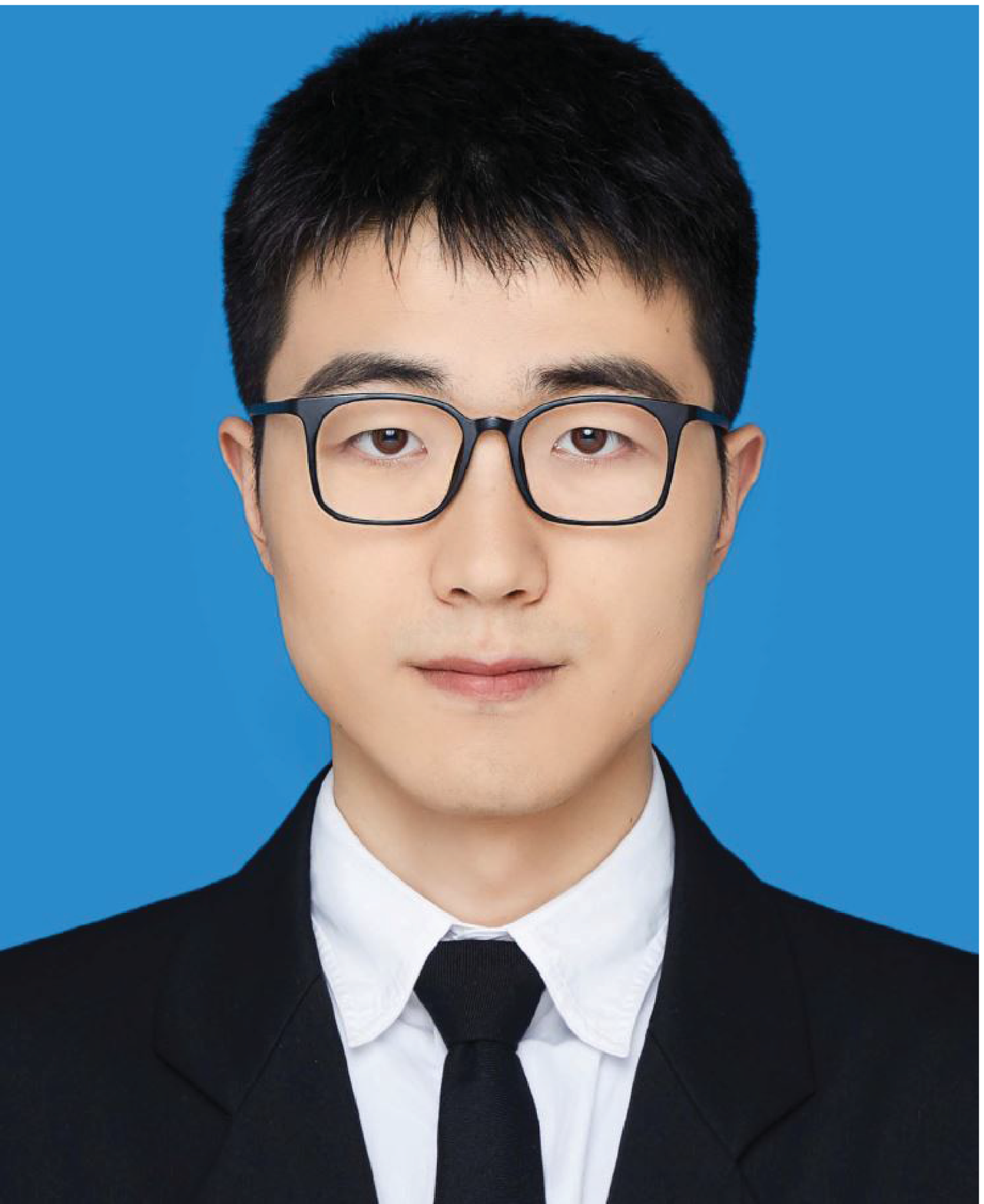}}]{Zhongbao Tian } received the B.E. degree in communication engineering from the University of Electronic Science and Technology of China (UESTC) in 2019. He is currently working toward the M.E. degree with the National Key Laboratory of Science and Technology on Communications, UESTC. He research and study interests include Terahertz communication and 3D beam forming.
\end{IEEEbiography}

\begin{IEEEbiography}
[{\includegraphics[width=1in,clip]{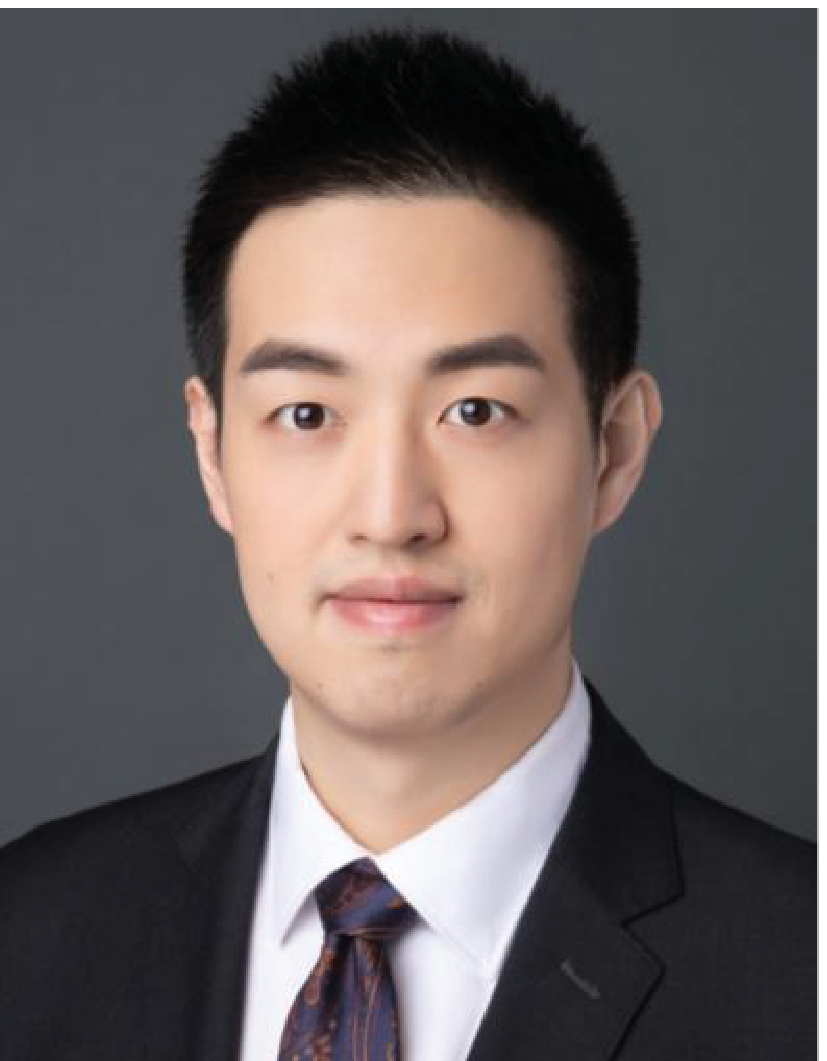}}]{Chong Han} received Ph.D. degree in Electrical and Computer Engineering from Georgia Institute of Technology, USA in 2016. He is currently an Associate Professor with the Terahertz Wireless Communications (TWC) Laboratory, Shanghai Jiao Tong University, China. He is the recipient of 2018 Elsevier NanoComNet (Nano Communication Network Journal) Young Investigator Award, 2017 Shanghai Sailing Program 2017, and 2018 Shanghai ChenGuang Program. He is an editor with IEEE Open Journal of Vehicular Technology since 2020, an associate editor with IEEE Access since 2017, an editor with Elsevier Nano Communication Network journal since 2016, and is a TPC chair to organize multiple IEEE and ACM conferences and workshops. His research interests include Terahertz communication networks, and electromagnetic nanonetworks. He is a member of the IEEE and ACM.

\end{IEEEbiography}

\begin{IEEEbiography}
[{\includegraphics[width=1in,clip]{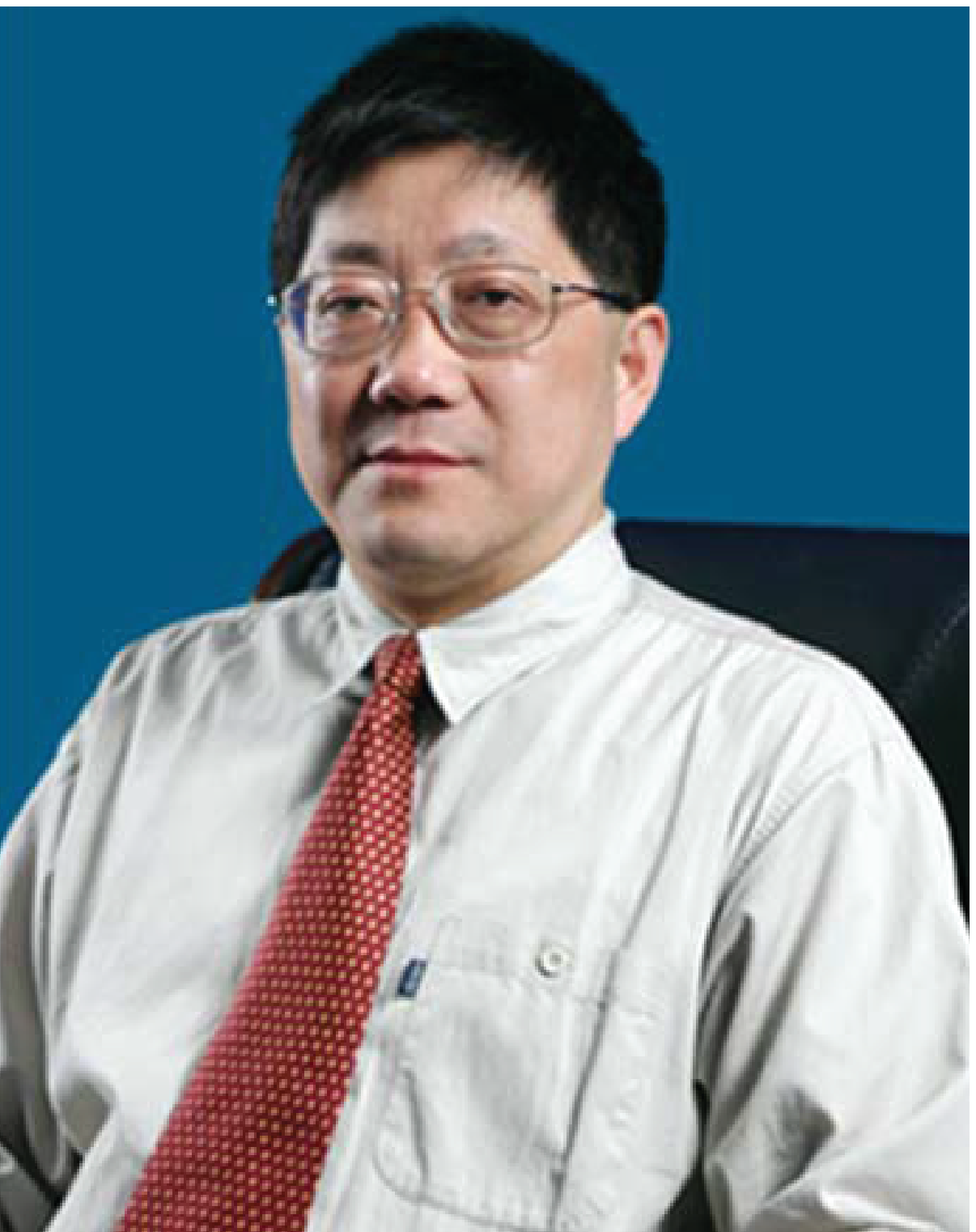}}]{Shaoqian Li} (Fellow, IEEE) received the B.E. degree in communication technology from Northwest Institute of Telecommunication Engineering (currently Xidian University), Xian, China, in 1981, and the M.E. degree in information and communication systems from the University of Electronic Science and Technology of China (UESTC), Chengdu, China, in 1984. He joined UESTC as an Academic Member in 1984, where he became a Professor of information and communication systems in 1997, and a Ph.D. Supervisor in 2000. He is currently the Director of the National Key Laboratory of Communications, UESTC. He has authored hundreds of journal or conference papers, and published several books. His research topics cover a broad range, including multiple-antenna signal processing technologies for mobile communications, cognitive radios, coding and modulation for next generation mobile broadband communications systems, wireless and mobile communications, anti-jamming technologies, and signal processing for communications. He has been a member of the Communication Expert Group of the National 863 Plan since 1998 and a member of The Future Project since 2005. He was the TPC Co-Chair of the IEEE International Conference on Communications, Circuits, and Systems in 2005, 2006, and 2008. He is currently a member of the Board of Communications and Information Systems of Academic Degrees Committee, State Council, China. He is also a member of the Editorial Board of the \emph{Chinese Science Bulletin} and the \emph{Chinese Journal of Radio Science}.
\end{IEEEbiography}

\end{document}